\documentclass[12pt, draftclsnofoot, onecolumn]{IEEEtran}
\normalsize

\usepackage{amssymb}
\usepackage{cite}

\ifCLASSINFOpdf
   \usepackage[pdftex]{graphicx}
   \DeclareGraphicsExtensions{.pdf,.jpeg,.png}
\else
\fi

\usepackage[cmex10]{amsmath}
\usepackage{algorithm,algorithmic}

\usepackage{array}
\newcolumntype{C}[1]{>{\centering\let\newline\\\arraybackslash\hspace{0pt}}m{#1}}

\usepackage[tight,footnotesize]{subfigure}

\usepackage{stfloats}
\usepackage{multirow}
\usepackage{booktabs}
\usepackage[flushleft]{threeparttable}
\usepackage{amssymb}
\usepackage{setspace}
\usepackage{stackrel}
\usepackage{wasysym}
\usepackage{enumerate}
\usepackage[export]{adjustbox}

\newtheorem{remark}{Remark}
\newtheorem{theorem}{Theorem}
\newtheorem{lemma}{Lemma}

\DeclareMathSizes{10}{8}{7}{6}
\IEEEaftertitletext{\vspace{-2\baselineskip}}

\begin{document}

\title{Analysis and design of Raptor codes using a multi-edge framework}
\author{Sachini~Jayasooriya,~\IEEEmembership{Student Member,~IEEE,}
	    Mahyar~Shirvanimoghaddam,~\IEEEmembership{Member,~IEEE,}    Lawrence~Ong,~\IEEEmembership{Member,~IEEE,}	   
        and~Sarah~J.~Johnson,~\IEEEmembership{Member,~IEEE}}

\maketitle
\doublespacing

\begin{abstract}
The focus of this paper is on the analysis  and design of   Raptor codes  using a multi-edge framework.  In  this regard, we first represent the Raptor code as a multi-edge type low-density parity-check (MET-LDPC) code. This MET representation  gives a  general framework to analyze and design  Raptor codes  over a binary input additive white Gaussian noise channel using  MET density evolution (MET-DE). We  consider a joint decoding  scheme based on the belief propagation (BP) decoding for Raptor codes in the multi-edge framework, and analyze the convergence behavior of the BP decoder using MET-DE.  In  joint decoding of Raptor codes,  the component codes correspond to  inner code and  precode are decoded in parallel and provide  information to each other. We also derive an exact expression for the stability of Raptor codes with joint decoding.   We then propose an efficient Raptor code design method using the multi-edge framework,  where we  simultaneously optimize  the inner code and the precode. Finally we consider  performance--complexity trade--offs of Raptor codes using the multi-edge framework.  Through density evolution analysis we show that the designed Raptor codes using the multi-edge framework outperform  the existing Raptor codes in literature  in terms of the realized rate.  
\end{abstract}

\begin{IEEEkeywords}
Belief-propagation, code optimization, density evolution, MET-LDPC codes, Raptor
codes  
\end{IEEEkeywords}

\section{Introduction}

The introduction of graph-based codes together with the belief propagation (BP) decoding~\cite{RichardsonIT2001capacity},  has made a significant change in the field of error correcting codes in terms of low-complexity  decoding. Generally, a graph-based code can be represented by a  Tanner graph in which  variable and check nodes respectively correspond to the codeword symbols and the parity check constraints~\cite{tannerIT1981recursive}.
One of the most prominent classes of graph-based codes for which the BP decoding algorithm gives near-capacity  performance  is  low-density parity-check (LDPC) codes invented by Gallager~\cite{GallagerTHESIS1963}. As a unifying framework for graph-based codes,  Richardson and Urbanke~\cite{RichardsonW2002multi} introduced multi-edge type low-density parity-check (MET-LDPC) codes to control the Tanner graph structure of random code ensembles. A code ensemble is the set of all possible  codes  with a particular degree distribution of their Tanner graph representation.

A numerical technique, called  MET density evolution (MET-DE)~\cite{richardsonBOOK2008modern} has been widely used for  analysis and design of MET-LDPC code ensembles under BP decoding. 
MET-DE determines the asymptotic behavior of the BP decoding for a given MET-LDPC code ensemble by iteratively tracking the  probability density function (PDF) of messages passed along the edges in the corresponding Tanner graph.  Applying MET-DE to a given MET-LDPC code ensemble, one can predict how codes (with a large enough code length) from that ensemble  behave on average.  MET-DE allows finding the optimal  degree distribution independent of the choice of code length or Tanner graph structure, which is the first step in any design or analysis of MET-LDPC codes.

Fountain codes~\cite{mackayIET2005fountain} are a class of graph-based codes, which have been inspired by the idea of \emph{rateless coding},  originally proposed for  data transmission over  erasure channels. Luby Transform (LT) codes~\cite{LubyISFCS2002LT} are the first class of efficient Fountain codes, which are essentially a type of low-density generator matrix (LDGM) codes~\cite{mackayIT1999good} with an average input bit degree that grows logarithmically.
This logarithmic growth causes several problems, such as increased complexity of  encoder and decoder, and also results in large error floors.   Raptor codes~\cite{shokrollahiIT2006raptor} are an extension of LT codes that overcome these problems. A Raptor code is a simple concatenation of an inner LT code with an outer code, called precode, which is usually a high rate LDPC code.  

Although the  Raptor code design has been well investigated for different binary channels~\cite{shokrollahiIT2006raptor,shirvanimoghaddamTcom2015raptor,hu2007TcomLperformance,etesamiIT2006raptor,chengTcom2009design,venkiahEURASIP2009jointly,barronMCC2009global,kuo2014design}, there has been little progress on universal design methods for  the binary-input additive white Gaussian noise (BI-AWGN) channel,  and a complete analysis of the asymptotic performance of  Raptor codes is still missing.  
Performance of Raptor codes over the BI-AWGN channel has been investigated in~\cite{etesamiIT2006raptor,chengTcom2009design}, and shown that the realized rate of a well-designed Raptor code can approach the capacity bound. Shirvanimoghaddam~\emph{et~al.}~\cite{shirvanimoghaddamTcom2015raptor}  studied the design of Raptor codes over BI-AWGN channels in the low signal-to-noise ratio (SNR) regime  and showed that properly designed Raptor codes can achieve rate efficiency larger than 0.95 in that regime.    Pakzad~\emph{et~al.}~\cite{pakzadITW2006design} studied  different design  principles, such as code length and complexity of the decoding, for finite-length Raptor codes under BP decoding for BI-AWGN channels.  
However, all these  methods  only considered the performance of  the LT code component of the Raptor code for a given precode.  A  high-rate regular LDPC code~\cite{venkiahEURASIP2009jointly,barronMCC2009global} or a high-rate LDPC code  with left-regular and right Poisson distributed parity check matrix~\cite{etesamiIT2006raptor,chengTcom2009design} has been widely used for the precode of Raptor codes in order to minimize the rate loss due to the precoding.  
This motivates us to propose a more general design framework for Raptor codes using a multi-edge framework. The multi-edge framework enables us to perform a comprehensive analysis of the asymptotic performance of the entire Raptor code including both the LT code and the LDPC code.

All the design methods proposed in literature on Raptor codes over BI-AWGN channel  have relied on linear programming based on approximate DE algorithms namely, the extrinsic information transfer (EXIT) chart analysis~\cite{ten2001convergence,chengTcom2009design} or Gaussian approximation~\cite{chungIT2001analysis}. These approximate DE algorithms are based on the assumption  of symmetric Gaussian distribution for the DE messages.   However, this assumption may not be accurate~\cite{jayasooriya2016new}, particularly at low rates and with punctured variable nodes, which is the case for Raptor codes. Therefore,  the  approximate DE algorithms can negatively impact the search for optimal code ensembles.  

The decoding process for Raptor codes generally consists of a series of decoding attempts, which progressively collects LT coded symbols until the decoder is confident that the transmitted message is correctly decoded. Basically  a Raptor decoding attempt consists of two decoding processes: the LT decoding process and the LDPC decoding process (collectively know as Raptor decoding process). Each of these decoding processes  runs a predetermined number of BP decoding iterations in which the soft information is passed back and forth along the edges in the corresponding Tanner graph.  The way that the soft information is exchanging  between two decoding processes  depends on the configuration of the Raptor decoder (i.e., tandem decoding or joint decoding).  The conventional design strategy of Raptor codes is mainly based on  tandem decoding, where the LT  component is decoded first and soft information is then send to the precoder. Venkian~\emph{et~al.}~\cite{venkiahEURASIP2009jointly} however showed that tandem decoding is sub-optimal compared to joint decoding, where both component codes are decoded in parallel and extrinsic information (i.e., the information from previous BP decoding iteration) is exchanged between the decoders.  Unfortunately, only few studies~\cite{venkiahEURASIP2009jointly} have considered joint decoding of Raptor codes. Further, there is no design strategy for the joint design of component codes of Raptor codes, using joint or tandem decoding.

Moreover, the traditional tandem decoding scheme proposed for Raptor codes starts the LT decoding process from scratch at every decoding attempt, without using soft information produced in the previous decoding attempt.  We refer to this decoding strategy as \emph{message-reset decoding}. Another possible decoding strategy for the LT decoding process in the  tandem decoding scheme is to  reuse the soft information produced in the previous decoding attempt. We refer to this decoding strategy as \emph{incremental decoding}. To-date only few results~\cite{hu2007TcomLperformance}  are available concerning the performance and the decoding complexity trade-off for Raptor codes over the BI-AWGN channel.
 
\subsection{Main contribution}
The main contributions of this paper are as follows: 
\begin{enumerate}
	\item  We study Raptor codes using the multi-edge framework, where we represent the Raptor code as a MET-LDPC code, which is the serial concatenation of an LT code  and an LDPC code (i.e., precode). This MET representation gives us a more unifying framework to analyze and design Raptor codes with the help of analytical tools used for MET-LDPC code, such as MET-DE~\cite{RichardsonW2002multi}.
	\item  We consider two decoding schemes for Raptor codes under the multi-edge framework, including tandem and joint decoding.  We derive the  stability conditions under tandem decoding using the multi-edge framework and re-derive  the existing stability conditions from~\cite{etesamiIT2006raptor}. We then derive an exact expression for the stability of  Raptor codes under joint decoding using the multi-edge framework. We  analyze Raptor codes using  both  tandem and joint decoding, and show the benefits of   joint decoding. 
	\item We then propose a joint Raptor code design method by including degree distributions of both LT and LDPC code components as variables  in the  optimization. This provides a framework for a better selection of the  LDPC code component depending on the channel SNR for which the code is designed (designed SNR) and the degree distributions of the LT code component.
	We formulate the optimization problem for Raptor codes for a given channel SNR as a non-liner cost function  maximization problem, where the  realized rate (which is computed from MET-DE) is the cost function and the degree distributions of LT and LDPC code components give the variables to be optimized. The realized rate of a Raptor code is defined as the ratio of the length of the information vector, and the average number of Raptor coded symbols required for the successful decoding at the destination.  Moreover, our optimization method is based on the MET-DE, which
	in turn helps to increase the accuracy of the final result in the code optimization. Because we are no longer use any approximate DE algorithms such as EXIT charts and Gaussian approximation, which are previously used to optimize Raptor codes. Through MET-DE, we show that  designed and/or evaluated Raptor codes using joint decoding always give better rate efficiency results than the one with tandem decoding. This seems to be significant when  the evaluated SNR is below  the designed SNR.
	\item Finally, we implement a tandem decoding scheme based on the incremental decoding strategy  using the multi-edge framework, propose a modification to reduce complexity and show the advantage of considering the incremental decoding strategy in the design of the Raptor codes.
\end{enumerate}

\subsection{Organization}
The rest of the paper is organized as follows. Section~\ref{sec-Raptor in MET} presents a representation of Raptor codes in the multi-edge framework. In Section~\ref{sec- Raptor codes decoding}, we describe two possible decoding schemes based on the BP decoding for Raptor codes using MET-DE. Section~\ref{sec-raptor design}  provides the design methods of Raptor codes in the multi-edge framework.  Analytical results are shown in Section~\ref{sec-Discussion}. Finally, Section~\ref{sec-conclusion} concludes the paper.

\section{Representation of  Raptor codes in the multi-edge framework} \label{sec-Raptor in MET}
\begin{figure}[!t]
	\centerline{
		\subfigure[An example Tanner graph of a Raptor code with parameters $(k,\mathcal{C}, \Omega(x))$ truncated at code length $m$, where $p_i$ represents the fraction of degree $i$ nodes.]{\includegraphics[width=0.6\linewidth]{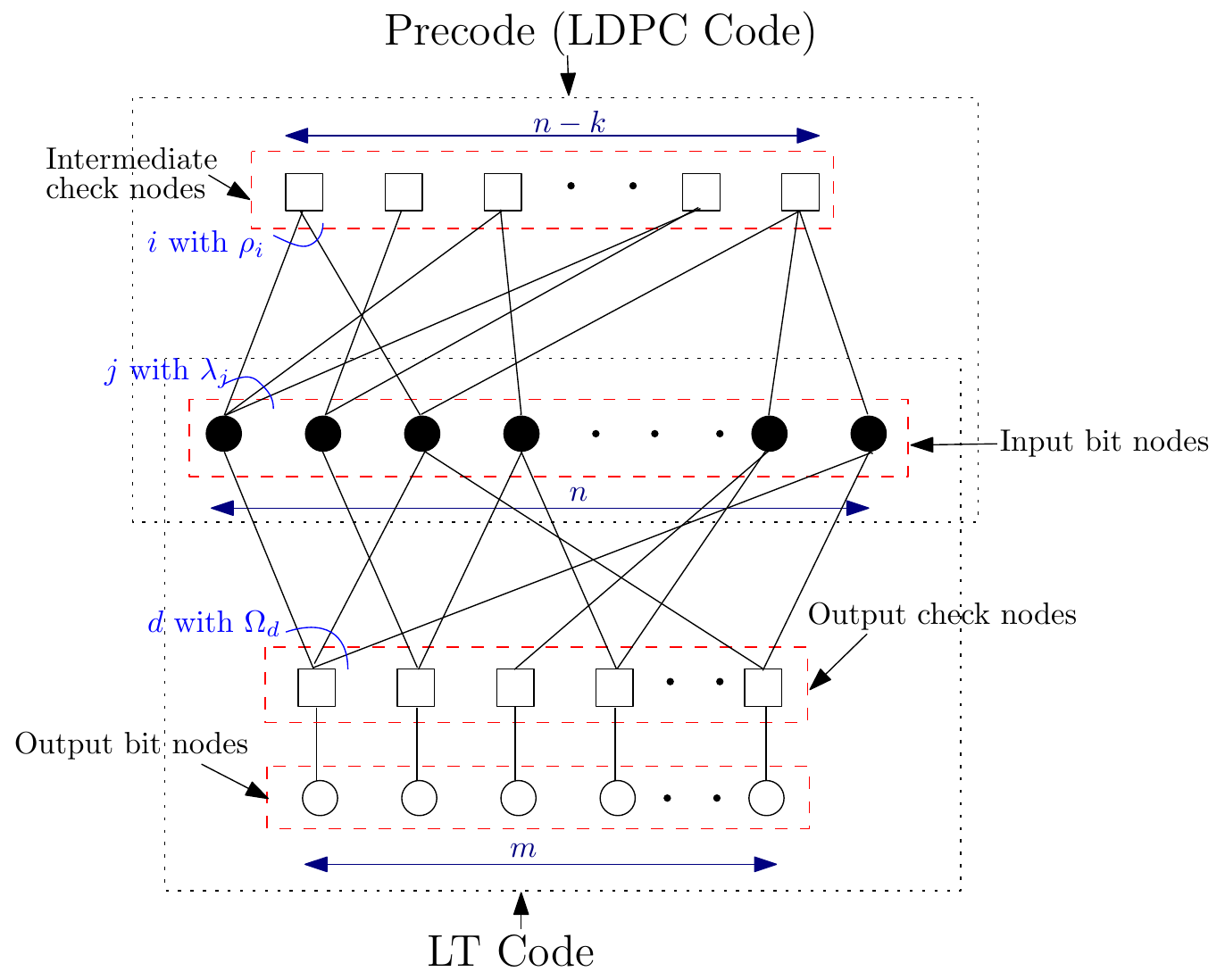}\label{fig:Raptor_code}}}	
		\vspace{0.5cm}
	\centerline{
		\subfigure[Graphical representation of a Raptor code ensemble as a MET-LDPC code ensemble, where    
	\textquoteleft$\Circle$\textquoteright, \textquoteleft$\CIRCLE$\textquoteright, and  \textquoteleft$\Square$\textquoteright~ respectively represent unpunctured variable nodes, punctured variable nodes and  check nodes, and \textquoteleft$\Pi_i$\textquoteright~ represents the edge-type $i$.	The number of nodes for different edge-types (i.e., $L_{[\dots]}$ and $R_{[\dots]}$) are defined as fractions of  $m$. Note that $k_v=k_c=1$.]{\includegraphics[width=0.8\linewidth]{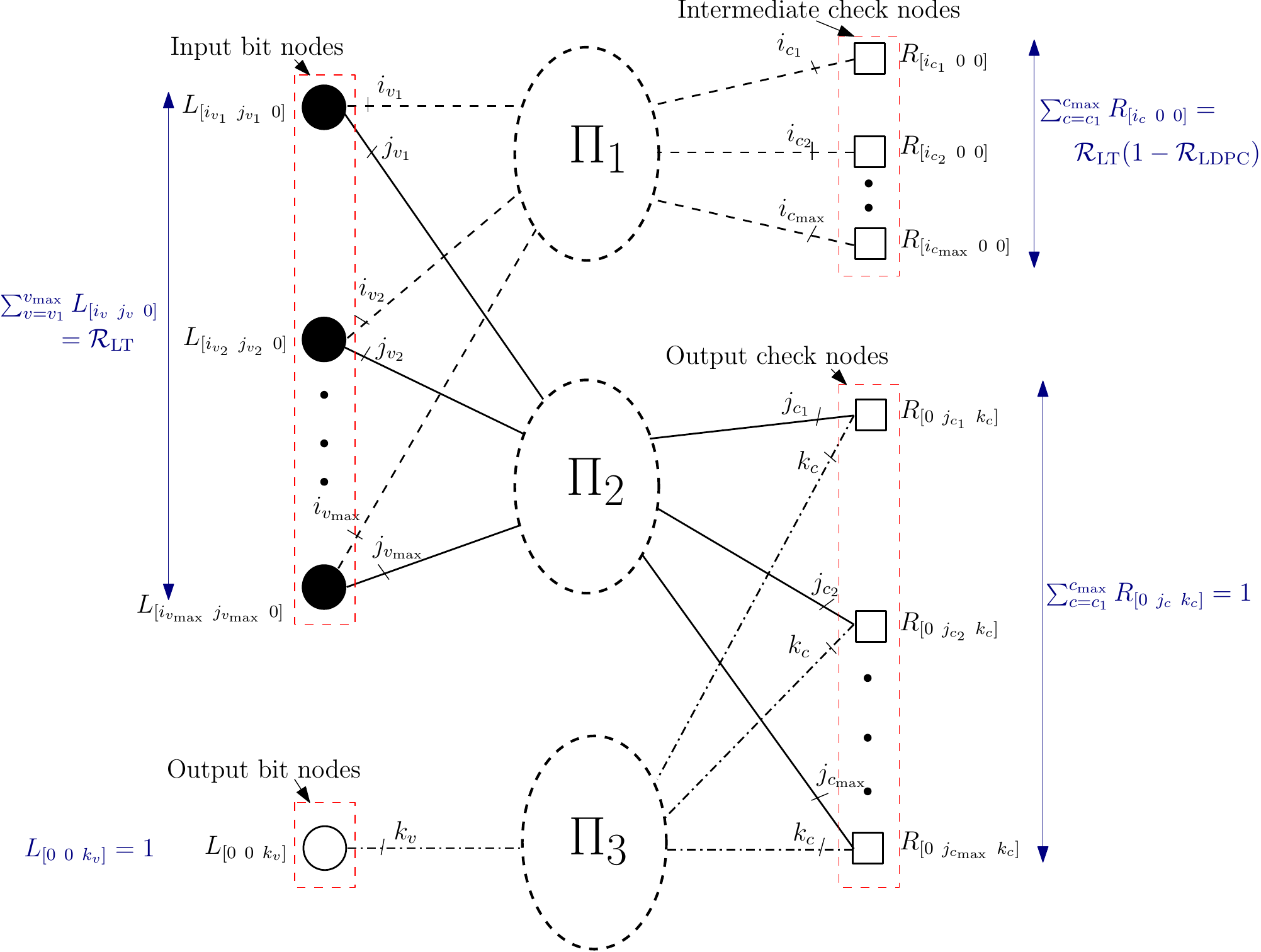}		\label{fig:Raptor_code_in_MET_detail}}}
	\caption{Graphical representation of a Raptor code.} 
\end{figure}

In this section, we describe the new representation of Raptor codes as  MET-LDPC codes. 
A Raptor code specified by parameters $(k,\mathcal{C},\Omega(x))$ is a serial concatenation of an inner LT code  with  degree distribution $\Omega(x)$,  and a precode $\mathcal{C}$, generally a high-rate LDPC code.  
For a Raptor code, first, the $k$ bit information vector is  precoded using a rate $\mathcal{R}_\mathrm{LDPC} = k/n$ LDPC code to generate $n$ LDPC coded symbols, referred to as input bits. Then using an LT code with degree distribution $\Omega(x)$, a potentially limitless number of LT coded symbols, referred to as output bits, are generated. Each output bit is the XOR of $d$ randomly selected input bits, where $d$ is randomly obtained from $\Omega(x)$. The realized rate of the LT code is $\mathcal{R}_\mathrm{LT} = n/\hat{m}$, where $\hat{m}$ is the average number of LT coded symbols required for a successful decoding. An example Tanner graph of a Raptor code truncated at code length $m$ is shown in Fig.~\ref{fig:Raptor_code}.
 
Using the multi-edge framework, the Tanner graph of a Raptor code ensemble can be drawn as shown in Fig.~\ref{fig:Raptor_code_in_MET_detail}. This  graph  has three edge-types, where edge-type $\Pi_1$ represents the precode   $\mathcal{C}$  and edge-type $\Pi_2$ represents the LT code with degree distribution $\Omega(x)$. Edge-type $\Pi_3$ is used to include the output bits which transfer  channel information to the LT code. Moreover, the sub-graph comprising with edge-types $\Pi_2$ and $\Pi_3$ can be considered as an LDGM code. 

The Raptor code ensemble  in the multi-edge framework can be represented by  two node-perspective multinomials associated with variable nodes and check nodes  as follows:  
\begin{align}
	\label{eq : raptor_varaible}
	L(\boldsymbol{r},\boldsymbol{x}) &=	r_0~\sum_{i_v=1}^{i_{v_{\max}}}~\sum_{j_v=1}^{{j_{v_{\max}}}}L_{[i_v~j_v~0]}~x_1^{i_v}~x_2^{j_v} ~+~ r_1~L_{[0~0~1]}~x_3, \\
	\label{eq : raptor_check}
	R(\boldsymbol{x}) &= \sum_{i_c=2}^{{i_{c_{\max}}}} R_{[i_c~0~0]}~x_1^{i_c} ~+~ \sum_{j_v=1}^{{j_{c_{\max}}}} R_{[0~j_c~1]}~x_2^{j_c}~x_3,	
\end{align}
where $\boldsymbol{x}$ and $\boldsymbol{r}$ are vectors defined as follows. The vector $\boldsymbol{x} = [x_1,x_2,x_3]$ corresponds to each edge-type in the Tanner graph and  $x_p^{k}$ is used to indicate the number of edges of the $p$th edge-type connected to  a particular  node. The vector $\boldsymbol{r} = [r_0,r_1]$ associated with each variable node, corresponds to the channel
to which the variable node is connected.  In the multi-edge framework, the  input bits are considered as punctured variable nodes (i.e., which are not transmitting over the channel) and  denoted  by  $r_0$.  The output bits are considered as unpunctured variable nodes (i.e., which are transmitting over a single channel) and denoted  by  $r_1$.   Finally, we use $[i~j~k]$ to categorize the node types, where $i, j$ and $k$, respectively denote the  number of edges of the edge-type 1, 2 and 3 connected to that node, i.e., $L_{[i_v~j_v~k_v]}$ and $R_{[i_c~j_c~k_c]}$ correspond to the fraction of variable nodes of type ${[i_v~j_v~k_v]}$ and the fraction of check nodes of type ${[i_c~j_c~k_c]}$, where the fractions are relative to the number of transmitted variable nodes.

In order to impose the Raptor code structure in the MET-LDPC framework, we add some additional constraints into (\ref{eq : raptor_varaible}) and (\ref{eq : raptor_check}). 
\begin{align}
L_{[0~0~1]} &= 1 \label{total_bits_1}, \\
\sum_{j_c=1}^{{j_{c_{\max}}}} R_{[0~j_c~1]} &= 1 \label{total_bits_2},\\
\sum_{i_v=2}^{i_{v_{\max}}}~\sum_{j_v=1}^{j_{v_{\max}}}L_{[i_v~j_v~0]}  &= \mathcal{R}_{\mathrm{LT}}, \label{LT}\\
\sum_{i_c=1}^{{i_{c_{\max}}}} R_{[i_c~0~0]} &= \mathcal{R}_{\mathrm{LT}}(1-\mathcal{R}_{\mathrm{LDPC}}).\label{LDPC}
\end{align}
Constraints  (\ref{total_bits_1}) and (\ref{total_bits_2}) are used to satisfy the constraints on the total number of transmitted bits as fractions of code length. Constraints (\ref{LT}) and (\ref{LDPC})  are used to compute the rates of LT  and LDPC  code components in multi-edge framework.  
The realized rate of the Raptor code for a given SNR, $\gamma$, in the multi-edge framework can be computed as~{\protect\cite[page 383]{richardsonBOOK2008modern}}
\begin{align}
\mathcal{R}(\gamma) &= L(\boldsymbol{1},\boldsymbol{1}) - R(\boldsymbol{1}),
\label{code rate}
\end{align} 
where $\boldsymbol{1}$ denotes a vector of all $1$'s with the length determined by the context. The   rate efficiency of the Raptor code is then computed as 
\begin{align}
\eta(\gamma) &= \frac{\mathcal{R}(\gamma)}{C(\gamma)},  \label{Rptor_rate_eff}
\end{align}
where $C(\gamma)$ is the capacity of the BI-AWGN channel at SNR, $\gamma$, which is given by~\cite{etesamiIT2006raptor} 
\begin{align}
C(\gamma) &= 1 - \frac{1}{\sqrt{8\pi \gamma}}\int_{-\infty}^{\infty} \log_2\left(1+e^{-x}\right) e^{-\frac{(x-2\gamma)^2}{8 \gamma}} dx .
\end{align}

Finally, we specify the  standard notation of the Raptor code with parameter $(k, \mathcal{C}, \Omega(x))$ using the MET notation as follows.  The node-perspective degree distribution  related with LT code, $\Omega(x)$, can be computed from (\ref{eq : raptor_check}) as follows: 
\begin{align}
	\Omega(x) &= \sum_{j_c=1}^{j_{c_{\max}}} R_{[0~j_c~1]}~x^{j_c},
	\label{omega_x}
\end{align}	
where $R_{[0~j_c~1]}$ denotes the fraction of output bits with degree $j_c$, which is  $\Omega_{j_c}$. 
The edge-perspective degree distributions associated  with variable nodes and check nodes of the precode ($\lambda(\boldsymbol{x})$ and $\rho(\boldsymbol{x})$) can be computed from (\ref{eq : raptor_varaible}) and (\ref{eq : raptor_check}) as follows:
\begin{align}
\lambda(\boldsymbol{x}) &= \frac{L_{x_1}(\boldsymbol{r},\boldsymbol{x})}{L_{x_1}(\boldsymbol{1},\boldsymbol{1})} \hspace*{1cm} \text{and}  \hspace*{1cm}
\rho(\boldsymbol{x}) = \frac{R_{x_1}(\boldsymbol{x})}{R_{x_1}(\boldsymbol{1})} \label{LDPC_dd},
\end{align}	
where $L_{x_1}(\boldsymbol{r},\boldsymbol{x}) = \frac{d L(\boldsymbol{r},\boldsymbol{x})}{d{x_1}} $ and $R_{x_1}(\boldsymbol{x}) = \frac{d R(\boldsymbol{x})}{d{x_1}} $.

\section{Decoding of  Raptor codes in the multi-edge framework}\label{sec- Raptor codes decoding}

As stated earlier, a Raptor decoding process generally proceeds with several decoding attempts, where in each  decoding attempt, Raptor decoding process runs a predetermined number of BP decoding iterations to obtain a message estimate. Specifically, the Raptor decoder begins the first decoding attempt after collecting $m_f$ number of received bits (i.e., the noise corrupted output bits). Then after the Raptor decoding process, it checks the errors in the  recovered message using CRC bits embedded during LDPC encoding.  If no error is found, i.e., the decoding attempt is successful, an acknowledgment is sent via a noiseless feedback channel and  terminates the Raptor decoding process. If the first decoding attempt is not successful, Raptor decoder collects another $\delta m$  received bits  and begins the second decoding attempt. This process is repeated until no 
error is found in the recovered message. In the Raptor decoding process,  the size of the Tanner graph for a particular decoding attempt  depends on the total  number of  output bits used in  that decoding attempt. For example,  at $p$th decoding attempt, the corresponding Tanner graph has $(n+m)$  variable nodes and $(n-k+m)$  check nodes, where $m=m_f + (p-1)\delta m$ (See Fig.~\ref{fig:Raptor_code}). 

In this section, we consider two possible configurations for Raptor decoding process in the multi-edge framework:   joint decoding scheme and  tandem decoding scheme.  Note that we use the message-reset decoding strategy throughout the paper unless otherwise stated.

\subsection{The joint decoding scheme based on the BP decoding and the MET-DE for Raptor codes}  \label{MET-DE}
In  joint decoding,  LT and LDPC  code components are decoded in parallel and provide  extrinsic information to each other. One of the main advantages of representing Raptor codes in a multi-edge framework is that we can easily  analyze  joint decoding scheme based on the  BP decoding using MET-DE.   

In the BP decoding, there are three types of messages passing along the edges of the corresponding Tanner graph, namely channel message, variable-to-check message and check-to-variable message. The channel message is categorized as an intrinsic message, whereas the variable-to-check message and the check-to-variable message are  categorized as extrinsic messages~{\protect\cite[pages 390-391]{ryanBook2009channel}}.  
Let $m_{\mathrm{v} \rightarrow \mathrm{c}}^{(\ell)}$   denote  the variable-to-check message from variable node $\mathrm{v}$ to check node $\mathrm{c}$ at the $\ell{\text{th}}$ iteration of the BP decoding,   $m_{\mathrm{c} \rightarrow \mathrm{v} }^{(\ell)}$ denote the  check-to-variable message from check node $\mathrm{c}$ to variable node $\mathrm{v}$ at the $\ell{\text{th}}$ iteration of the BP decoding, and $m_0$ denote the channel message.
For the $\ell{\text{th}}$ BP decoding iteration, $m_{\mathrm{v} \rightarrow \mathrm{c}}^{(\ell)}$ and $m_{\mathrm{c} \rightarrow \mathrm{v} }^{(\ell)}$  can be computed as follows: 
\begin{align}
m_{\mathrm{v} \rightarrow \mathrm{c}}^{(\ell)} =
\begin{cases}
m_0 & \text{if~} \ell=1,\\
m_0 + \sum_{\mathrm{c}'\in C_{\mathrm{v}}\backslash \mathrm{c}}m_{\mathrm{c}' \rightarrow \mathrm{v} }^{(\ell-1)} & \text{if~} \ell>1,
\end{cases}
\label{v-to-c msg}
\end{align}
\begin{align}
m_{\mathrm{c} \rightarrow \mathrm{v} }^{(\ell)} &= 2 \tanh^{-1} \left(\prod_{\mathrm{v}'\in V_{\mathrm{c}}\backslash \mathrm{v}}\tanh\left(\frac{m_{\mathrm{v}' \rightarrow \mathrm{c}}^{(\ell)}}{2}\right) \right), \label{c-to-v msg}
\end{align}
where $C_{\mathrm{v}}\backslash \mathrm{c}$ is the set of check nodes connected  to variable node $\mathrm{v}$ excluding check node $\mathrm{c}$, and  $V_{\mathrm{c}}\backslash \mathrm{v}$ is the set of variable nodes connected  to check node $\mathrm{c}$ excluding variable node $\mathrm{v}$.

For the BP decoding on a BI-AWGN channel,  intrinsic and extrinsic messages  are described by PDFs for analysis using MET-DE.   Let $f\big({m_{\mathrm{v} \rightarrow \mathrm{c}}^{(\ell)}}\big)$ and $f\big({m_{\mathrm{c} \rightarrow \mathrm{v}}^{(\ell)}}\big)$, respectively denote the PDF of the  message from variable node $\mathrm{v}$ to check node $\mathrm{c}$ and, the PDF of the  message from check node $\mathrm{c}$ to variable node $\mathrm{v}$,  at the $\ell{\text{th}}$ BP decoding iteration. Let $f\big(m_{\mathrm{0}}\big)$ denote the PDF of the channel message.
Then from~(\ref{v-to-c msg}) and~(\ref{c-to-v msg}), the updated message PDFs for variable nodes and check nodes at the $\ell$th BP decoding iteration can be computed as follows:
\begin{align}
\label{eq:VN to CN PDF MET}
f\big({m^{(\ell)}_{\mathrm{v} \rightarrow \mathrm{c}}}\big) &=
\begin{cases}
	f\big(m_{0}\big) & \text{if~} \ell=1,\\
	f\big(m_{0}\big) \otimes \left[ \stackrel[\mathrm{c}'\in C_{\mathrm{v}}\backslash \mathrm{c}]{}{\otimes} f\big({m^{(\ell-1)}_{\mathrm{c}' \rightarrow \mathrm{v} }}\big)\right] & \text{if~} \ell>1,
\end{cases} \\
\label{eq:CN to VN PDF MET}
f\big({m^{(\ell)}_{\mathrm{c} \rightarrow \mathrm{v}}}\big) &=  \stackrel[\mathrm{v}'\in V_{\mathrm{c}}\backslash \mathrm{v}]{}{\boxtimes}  f\big({m^{(\ell)}_{\mathrm{v}' \rightarrow \mathrm{c}}}\big), 
\end{align}
where $\otimes$ denotes the variable node convolution and $\boxtimes$ denotes the check node convolution. For more details we refer readers to Richardson and Urbanke~{\protect\cite[pages 390-391 and 459-478]{richardsonBOOK2008modern}}.

\vspace*{-1em}

\subsection{Stability of Raptor codes with joint decoding  using the multi-edge framework} 
The stability analysis using MET-DE examines the asymptotic behavior of the BP decoding when it is close to a successful decoding and gives a sufficient condition for the convergence of  the bit error rate (BER)  to zero  as the BP decoding iteration, $\ell$, tends to infinity. Before analyzing the stability condition with the joint decoding scheme, we refer readers to Remark~\ref{stability} and Lemma~\ref{Degree-one} given in Appendix~\ref{appendix_A}, upon which our analysis is based.

Now consider  Raptor codes  represented in a multi-edge framework. As shown in Fig.~\ref{fig:Raptor_code_in_MET_detail}, the
MET representation of a Raptor code has degree-one variable nodes. Therefore, messages on edge-types which are connected to degree-one variable nodes may not converge to zero BER even thought output BER converges to zero. Thus we need to consider the special case which is given in Remark~\ref{stability}. Recall that in the MET setting, a Raptor code is a serial concatenation of an LT code  with an LDPC code.  According to Remark~\ref{stability}, we can categorize edge-types in Fig.~\ref{fig:Raptor_code_in_MET_detail} as $E_1=\{\Pi_1\}, ~E_2=\{\Pi_3\}$  and $E_{1,2}=\{\Pi_2\}$ and define nodes connected to $\Pi_1$, i.e., the LDPC code component of the Raptor code,  as the core LDPC graph for the stability analysis. The edge-perspective multinomial of the  core LDPC graph with edges connected to $\Pi_2$ is  given by 
\begin{align}
\lambda_1(x_1, x_2) &=\sum_{i\geq2} \sum_{j\geq 1} \lambda_{[i~j]}~x_1^{i-1}~x_2^{j}, \label{core_VN} \\
\rho_1(x_1) &= \sum_{i\geq 0} \rho_{i}~x_1^{i-1}, \label{core_CN}
\end{align}
where we assume that edges from $\Pi_2$, all carrying a fixed message PDF  in check-to-variable direction.  Since the stability analysis considers the termination of the BP decoding, i.e., $\ell \rightarrow \infty$, from Lemma~\ref{Degree-one}, we can assume that edges from $\Pi_2$, all carry a  message with the same distribution as the channel message in the check-to-variable direction. Then the stability condition for Raptor codes with joint decoding is given in Theorem~\ref{Stability_Raptor}.

\begin{theorem}[Sufficient condition for the stability of Raptor codes decoded with joint decoding  based on the BP decoding]\label{Stability_Raptor}
Consider a Raptor code decoded with a joint decoding  based on the BP decoding  using the multi-edge framework.
On a BI-AWGN channel, the  stability condition is given by	
	\begin{align*}			 
	\sum_{j\geq 1} \lambda_{[2~j]}(x_0)^j~\rho_1'(1) &< 1 ,
	\end{align*}
	where $\lambda_{[2~j]}$ gives the fraction of degree-two variable nodes in the LDPC code component and $x_0$  is the Bhattacharyya constant~{\protect\cite[pages 202]{richardsonBOOK2008modern}} associated with the channel message with noise variance $\sigma^2$, and $\rho_1'(1) = \frac{d\rho_1(x_1)}{d x_1}\big|_{x_1=1}$. 
\end{theorem}

\begin{IEEEproof}
	See Appendix~\ref{appendix_B}	
\end{IEEEproof}
Note that Theorem~\ref{Stability_Raptor}  gives an upper bound on the fraction of degree-two variable nodes in the LDPC code component, which is similar to the stability condition of  standard LDPC codes  having multiple channel inputs.

\vspace*{-1em}

\subsection{The tandem decoding scheme based on the BP decoding for Raptor codes using the multi-edge framework} 
Following the same procedure as for  joint decoding, we can easily implement and analyze   tandem decoding in  the multi-edge framework  using equations (\ref{v-to-c msg}) to (\ref{eq:CN to VN PDF MET}). The main difference between joint decoding and tandem decoding is that, in tandem decoding, LT and LDPC code components are  decoded independently.  That is at the first stage, we apply DE equations to the Tanner graph structure shown in Fig.~\ref{fig:Raptor_code_in_MET_detail} assuming that there are no messages coming from edge-type $\Pi_1$.   Subsequently, once the predetermined criteria for the first stage  is satisfied (such as the target minimum  mean LLR ($\mu_0$) or the maximum number of BP decoding iterations for the LT code component), the decoded LLR of each input bit is computed as,   
\begin{align}
\mathcal{L}_{\mathrm{v}} = \sum_{\mathrm{c} \in C_{\mathrm{v}}} m_{\mathrm{c}\rightarrow \mathrm{v}}^{(\ell)},  \label{decoded_LLR}
\end{align}
where $\mathcal{L}_{\mathrm{v}}$ denotes the decoded LLR  of input bit $\mathrm{v}$. At the second stage, we apply DE equations to the Tanner graph structure shown in Fig.~\ref{fig:Raptor_code_in_MET_detail} assuming that the check-to-variable messages coming from edge-type $\Pi_2$ are fixed and equal to the decoded LLRs.

\subsection{Stability of Raptor codes with tandem decoding  using the multi-edge framework} 
In the stability analysis of tandem decoding, we generally examine the asymptotic behavior of the BP decoder  at the beginning of  the LT decoding process, assuming that the decoding of LDPC codes is successful. Thus the stability condition of Raptor codes with tandem decoding  gives a condition to successfully start  the BP decoding rather than the decoding convergence as in  joint decoding. In this section, we  validate the stability condition given in~\cite{etesamiIT2006raptor} for tandem decoding using the multi-edge framework. 

The edge-perspective multinomial related to the LT code can be computed in the multi-edge framework as follows: 
\begin{align}
	\lambda_2(x_2) &= \frac{L_{x_2}(\boldsymbol{r}, \boldsymbol{x})}{L_{x_2}(\boldsymbol{1}, \boldsymbol{1})} = e^{\alpha(x_2-1)},		 \label{tandem_lamda}\\ 
	\rho_2(x_2) &= \frac{R_{x_2}(\boldsymbol{x})}{R_{x_2}(\boldsymbol{1})} = \sum_i \left( \frac{i~ \Omega_i}{\sum_i  i~ \Omega_i} \right) x_2^{i-1} ,\label{tandem_rho}
\end{align}
where $\Omega_i$ can be computed from (\ref{omega_x}).  Etesami~\cite{etesamiIT2006raptor} showed that the input bit degree distribution is binomial and can be approximated by a  Poisson distribution with parameter $\alpha$. Thus  $\lambda_2(x_2)$ is a  Poisson distribution, where $\alpha$ is the average input bit degree.  Then the stability condition for Raptor codes with tandem decoding is given in Theorem~\ref{Stability_Raptor_tandem}.
\begin{theorem}[Stability of Raptor codes decoded with tandem decoding  based on the BP decoding]\mbox{}\label{Stability_Raptor_tandem}
Consider a Raptor code decoded with tandem decoding  based on the BP decoding  using the multi-edge framework. Let  $\Omega_1$ and $\Omega_2$, respectively denote the fractions of output bits with degree-one and degree-two, and $\alpha$ and $\beta$, respectively give the average degree of input bits and output bits.  
Then on a BI-AWGN channel, the BP decoding process can be started successfully,  if			 
$	\Omega_1 \geq 0$   and
$\Omega_2 \geq {\beta}/{2 \alpha \tilde{x_0}}$, where $\tilde{x_0}$  is the D-mean~{\protect\cite[pages 201]{richardsonBOOK2008modern}} associated with a channel message with noise variance $\sigma^2$.
Moreover, for a capacity approaching code we have  $\Omega_2 \geq {C(\gamma)}/{2  \tilde{x_0}}$, where $C(\gamma)$ is the capacity of the BI-AWGN channel with SNR, $\gamma$.
\end{theorem}
\begin{IEEEproof}
	See Appendix~\ref{appendix_C}	
\end{IEEEproof}

To be more comprehensive, we also derive  the stability condition for the LDPC code component of the Raptor code using MET-DE. In this case, we examine the asymptotic behavior of the BP decoding when it is close to a successful  decoding, i.e.,  the BER of input bit nodes converge  to zero.
Recall that the edge-perspective degree distributions  associated with the LDPC code is denoted by $\lambda(x)$ and $\rho(x)$. Under tandem decoding, LDPC code is decoded using  decoded LLRs  computed for each input bit node at the end of the LT decoding process. In the stability analysis we consider these decoded LLRs  as channel inputs to each input bit node. Then the stability condition for LDPC code component of the Raptor code with tandem decoding is given in Theorem~\ref{Stability_Raptor_LDPC}.    

\begin{theorem}[Sufficient condition for the stability of the LDPC code component of the Raptor code decoded with tandem decoding based on the BP decoding ]\mbox{}\label{Stability_Raptor_LDPC}
	Consider a Raptor code decoded with tandem decoding  based on the BP decoding  using the multi-edge framework.
	On a BI-AWGN channel, the decoding of the LDPC code component of the Raptor code is successful if	 
$	\lambda_{2}~\rho'(1) \leq 1 / \mathcal{B}(\mathcal{L}_2)$,
	where $\lambda_{2}$ gives the fraction of degree-two variable nodes (i.e., degree-two input bit nodes) in the LDPC code component and $\mathcal{B}(\mathcal{L}_2)$  is the Bhattacharyya constant~{\protect\cite[pages 202]{richardsonBOOK2008modern}} associated with the decoded LLR of input bit nodes with degree-two. 
\end{theorem}

\begin{IEEEproof}
	We can proceed the same way as with Theorem~\ref{Stability_Raptor} to complete the proof.
\end{IEEEproof}

\vspace*{-1em}

\subsection{Tandem decoding scheme based on the incremental decoding strategy} 

\begin{figure}[!t]
	\centerline{
		\subfigure[An example Tanner graph of the LT code component of a Raptor code truncated at code length $m+\delta m$.]{\includegraphics[width=0.5\linewidth,valign=c]{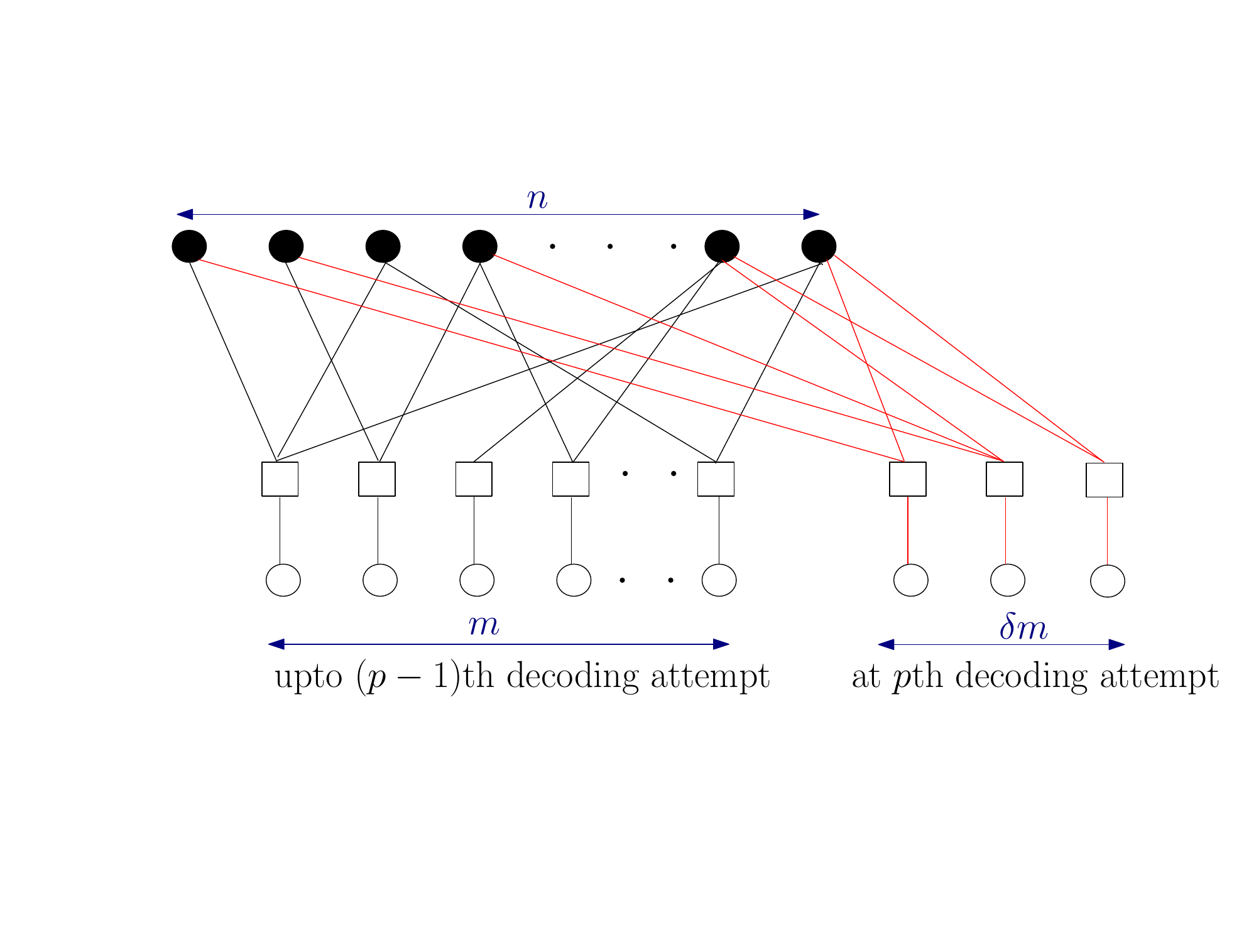}}	
		\hfill
		\subfigure[The  Tanner graph  of the LT code component of the Raptor code ensemble as a MET-LDPC code ensemble for incremental decoding strategy. ]{\includegraphics[width=0.5\linewidth,valign=c]{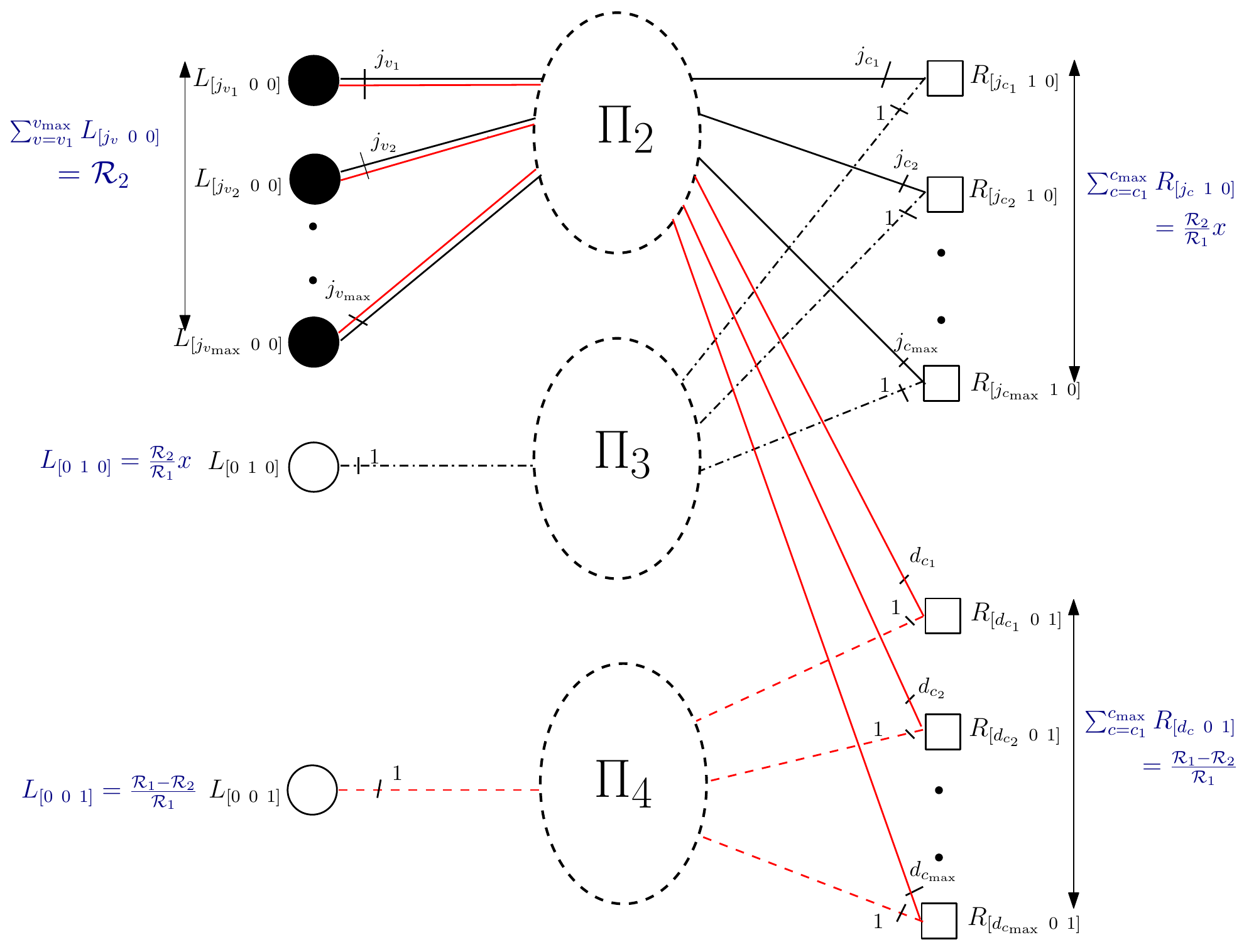}	}}
	\caption{Graphical representation  of the LT code component of the Raptor code ensemble at $p$th decoding attempt with incremental decoding strategy using the multi-edge framework.  $x$ denotes the fraction of output bits used from previous decoding attempts. $\mathcal{R}_1$ and $\mathcal{R}_2$ respectively denote  realized rates at the $(p-1)$th decoding attempt and the $p$th decoding attempt. Not shown are the ($1-x$) fraction of output bit nodes not used in this decoding round.} 
	\label{fig:figraptorcodeincremetal}
\end{figure}

In this section, we  implement the  incremental decoding strategy for tandem decoding  using the multi-edge framework.  The difference between the message-reset decoding strategy and the  incremental decoding strategy is  the way that they initialize the BP decoding algorithm  for the LT decoding process.  That is, with the incremental decoding strategy, the initial variable-to-check messages (i.e., when $\ell=1$) passed in the LT decoding process  at the $p$th decoding attempt  is set to the corresponding   messages computed at the end of the $(p-1)$th decoding attempt~\cite{hu2007TcomLperformance}, whereas with the message-reset decoding strategy these  initial variable-to-check messages are initialized to zero at every decoding attempt.  Under the incremental decoding strategy, the initial variable-to-check messages passed in the LT decoding process  at the $p$th decoding attempt ($m_{\mathrm{v} \rightarrow \mathrm{c}}^{(\ell=1)}$) can be computed as follows: 
\begin{align}
m_{\mathrm{v} \rightarrow \mathrm{c}}^{(\ell =1)} =
\begin{cases}
m_0 & \text{\small for a new variable node} ,\\
m_0 + \sum_{\mathrm{c}'\in C_{\mathrm{v}}\backslash\mathrm{c}} m_{\mathrm{c}'  \rightarrow \mathrm{v} }^{(T)} & \text{\small for a variable node used in previous decoding attempt} ,
\end{cases}
\label{v-to-c msg inc}
\end{align}
where $m_{\mathrm{c}' \rightarrow \mathrm{v} }^{(T)}$ denotes the check-to-variable message computed at the end of the $(p-1)$th decoding attempt.   
Moreover in our implementation, we propose a modification to incremental decoding strategy, where we only use  a fraction of output bits received during the previous decoding attempts  in order to further minimize the decoding complexity of the incremental decoding strategy. As shown in  Fig.~\ref{fig:figraptorcodeincremetal}, we implement the incremental decoding strategy using three edge-types (for the LT code component), where the $m$ output bits received during previous decoding  attempts are connected to edge-type $\Pi_3$, and the $\delta m$ new output bits received at the current  decoding attempt are connected to edge-type  $\Pi_4$. We  use $x$ to  denote the fraction   of output bits used from the previous decoding attempts. The parameter $x$ can be used as a controlling parameter to trade--off between performance and decoding complexity and $x$ can  take values from  0 to 1, where $x=1$ corresponds to  traditional incremental decoding.

\section{Design of  Raptor codes in the MET framework} \label{sec-raptor design}

\subsection{Problem statement for Raptor code optimization in the multi-edge framework  }
\label{sec:optimzation sta MET}
For the Raptor code optimization, we consider a Raptor code as a MET-LDPC code, and  formulate the optimization problem to  maximize  the design rate, $\mathcal{R}(\gamma)$, for a given channel SNR, $\gamma$, and maximum check node degrees, $i_{c_{\max}}$ and $j_{c_{\max}}$, such that  the BER of variable nodes decreases through BP decoding iterations. One of the main features of the proposed optimization method is that we jointly optimize the LT and the LDPC code components. We can formulate the optimization problem for the Raptor code in the multi-edge framework using joint decoding scheme as follows:

For a given SNR, $\gamma$, and maximum check node degrees, $i_{c_{\max}}$ and $j_{c_{\max}}$,
\begin{align}
\begin{aligned}
& {\mathrm{Maximize}} 
& & \mathcal{R}(\gamma) \\
& \mathrm{Subject~to}
& & (i)~ L_{[0~0~1]} = 1, \\
&&& (ii)~ \sum_{j_c=1}^{j_{c_{\max}}} R_{[0~j_c~1]} = 1, \\
&&& (iii)~ L(\boldsymbol{1},\boldsymbol{1}) - R(\boldsymbol{1}) = \mathcal{R}(\gamma), \\
&&& (iv)~ L_{x_i}(\boldsymbol{1},\boldsymbol{1}) = R_{x_i}(\boldsymbol{1}),\\
&&& (v)~ \max_j\{P_j\} < P^*, 
\label{optimization_joint}
\end{aligned}
\end{align}
where   $P_j$ and $P^*$ respectively denote  the BER  on the $j$th variable node and target maximum BER.  Here  $\mathcal{R}(\gamma)$ can be determined using MET-DE. Note that constraints $(i)$ and $(ii)$ were described in (\ref{total_bits_1}) and (\ref{total_bits_2}), constraint $(iii)$ imposes the code rate, constraint $(iv)$ is  to  satisfy  the total  number of edges of each edge-type in variable node side and check node side. This is known as  the socket count equality constraint, where
\begin{align*}
L_{x_i}(\boldsymbol{1},\boldsymbol{1}) = \frac{d L(\boldsymbol{r},\boldsymbol{x})}{d{x_i}} \biggr|_{\boldsymbol{r}=\boldsymbol{1},\boldsymbol{x}=\boldsymbol{1}} \hspace*{1cm} \text{and}  \hspace*{1cm}
R_{x_i}(\boldsymbol{1}) = \frac{d R(\boldsymbol{x})}{d{x_i}} \biggr|_{\boldsymbol{x}=\boldsymbol{1}}. 
\end{align*}
Finally, constraint $(v)$ is  to make sure that the MET-DE returns a  smaller BER than  the predefined BER value.

\subsection{Problem statement for Raptor code optimization for fixed precode settings}\label{sec:problem stat for ficed precode}
In the majority  of  literature concerning Raptor code design, the precode is fixed in advance (usually a high-rate regular or irregular LDPC code) and only the LT code is optimized for a given precode setting, using linear programming based on approximate DE algorithms.
Therefore, for a fair comparison with the existing results in  literature, we reformulate the optimization problem given in (\ref{optimization_joint}), where we optimize the Raptor code using the multi-edge framework and the  MET-DE  to   maximize the design rate of the LT code, $\mathcal{R}_{\mathrm{LT}}$ (i.e., only $L_{x_2}$ and $R_{x_2}$ are optimized).
This  is equivalent  to maximizing the design rate of Raptor code $\mathcal{R}(\gamma)$ for a  given LDPC code rate, $\mathcal{R}_{\mathrm{LDPC}}$.

The reformulated  optimization problem for the Raptor code with fixed precode settings in the multi-edge framework using joint decoding scheme is as follows:

For a given SNR, $\gamma$, maximum check node degree, $j_{c_{\max}}$, and LDPC code rate, $\mathcal{R}_{\mathrm{LDPC}}$,
\begin{align}
\begin{aligned}
& {\mathrm{Maximize}} 
& & \mathcal{R}_{\mathrm{LT}} \\
& \mathrm{Subject~to}
& & (i)~ L_{[0~0~1]} = 1, \\
&&& (ii)~ \sum_{j_c=1}^{j_{c_{\max}}} R_{[0~j_c~1]} = 1,\\
&&& (iii)~ \sum_{i_v=1}^{i_{v_{\max}}}~\sum_{j_v=1}^{j_{v_{\max}}}L_{[i_v~j_v~0]}  = \mathcal{R}_{\mathrm{LT}}, \\
&&& (iv)~\sum_{i_c=1}^{{i_{c_{\max}}}} R_{[i_c~0~0]} = \mathcal{R}_{\mathrm{LT}}(1-\mathcal{R}_{\mathrm{LDPC}}),\\
&&& (v)~ L_{x_2}(\boldsymbol{1},\boldsymbol{1}) = R_{x_2}(\boldsymbol{1}), \\
&&& (vi)~ \max_j\{P_j\} < P^*.
\label{optimization_tandem}
\end{aligned}
\end{align}
Here $\mathcal{R}_{\mathrm{LT}}$ can be determined using MET-DE and  constraints $(iii)$ and $(iv)$  described in (\ref{LT}) and (\ref{LDPC}), which are used to impose the rates of LT and LDPC codes into the multi-edge framework. Note that if we use the tandem decoding scheme, we need an additional constraint on the $\mu_0$ (or maximum number of BP decoding  iterations in the LT decoder) to determine when the BP decoding process will be switched from LT decoder to LDPC decoder.

\subsection{Optimization of the  degree distribution in the multi-edge framework} \label{check node} 

Recall that  the degree distributions of the Raptor code ensemble in the multi-edge framework are denoted by $L(\boldsymbol{r},\boldsymbol{x})$ and $R(\boldsymbol{x})$.  
As stated in Sections~\ref{sec:optimzation sta MET} and~\ref{sec:problem stat for ficed precode},  our goal in the Raptor code optimization is to optimally choose the elements in $L(\boldsymbol{r},\boldsymbol{x})$ and $R(\boldsymbol{x})$
so that the corresponding code ensemble yields the largest possible realized rate, which can
be determined via MET-DE. 
We employ a combined optimization method proposed in~\cite{jayasooriya2016Joint} to
optimize MET-LDPC code ensembles to that  Raptor code ensembles which are represented in the multi-edge framework. The combined optimization method optimizes the node fractions (denoted by $L_{[i_v~j_v~k_v]}$ and $R_{[i_c~j_c~k_c]}$ in (\ref{eq : raptor_varaible}) and (\ref{eq : raptor_check}) ) and the node degrees (denoted by $[i_v~j_v~k_v]$ and $[i_c~j_c~k_c]$ in (\ref{eq : raptor_varaible}) and (\ref{eq : raptor_check}))  in parallel using two different optimization techniques.  In this work, we use adaptive range method~\cite{jayasooriya2016Joint} to optimize the node fractions and differential evolution~\cite{storn-book1997-differential}  to optimize the node degrees.

\subsubsection{Input bit node degree distribution}
The aim of the  Raptor code optimization in multi-edge framework is  to optimally choose  $L(\boldsymbol{r},\boldsymbol{x})$ and $R(\boldsymbol{x})$
so that the corresponding code ensemble yields the largest possible realized rate. Under multi-edge framework, there are two essential constraints that needs to be satisfied in order to have a valid Raptor code degree distribution, namely the code rate constraint and the socket count equality constraint. These two constraints allow us to include $L(\boldsymbol{r},\boldsymbol{x})$ (i.e., $L_{[i_v~j_v~k_v]}$ and $[i_v~j_v~k_v]$) as  dependent variables in the code optimization.   In our design method, we choose two or three non-zero variable node  degrees (which may be chosen consecutively) for the  $L(\boldsymbol{r},\boldsymbol{x})$, based on $R(\boldsymbol{x})$ in order to achieve the desired code rate. 
Because, as shown by Hussain~\emph{et~al.}~\cite{hussain2013regularized},
we have also observed that there is a  little difference in performance whether $L(\boldsymbol{r},\boldsymbol{x})$ is  almost-regular or Poisson. At the same time, selecting an almost regular input bit degree distribution significantly helps to minimize the computational complexity of MET-DE, which in turn helps for an efficient code optimization.  

\subsection{A remark on Raptor code design}

\subsubsection{Selecting $\mu_0$ for tandem decoding}
In the majority  of  literature concerning Raptor code design, which use  tandem decoding  to formulate the optimization problem,   $\mu_0$ plays an important role. However  no serious  effort  has been taken for  deriving the optimal choice of $\mu_0$  and it remains mostly heuristic.  We have investigated the optimal choice of $\mu_0$ using the multi-edge framework and observed that  the choice of $\mu_0$  depends on the rate and the structure of the  LDPC code selected for the precode and  the channel SNR.  Here we explain the reasons  behind this. 

There are two important  requirements  that needs to be satisfied when we are selecting a value for $\mu_0$. The first requirement is that the value of $\mu_0$ needs to be selected based on the target residual error which is expected to be corrected by the LDPC decoder. The second requirement is that  $ \mu_0 \leq \min (\mathcal{L}_\mathrm{v})$  in order to ensure a successful decoding at the LDPC decoder.  The first requirement concludes that the  choice of  $\mu_0$ depends on the  rate and
structure of the LDPC code as   the  error correcting performance of a code is    approximately determined by these two parameters.  

We then use Lemma~\ref{Degree-one}  in Appendix~\ref{appendix_A} to explain the second requirement. As shown in  Fig.~\ref{fig:Raptor_code_in_MET_detail}, the LT code component in MET setting has a set of degree-one variable nodes. Therefore according to Lemma~\ref{Degree-one}, after a sufficient number of BP decoding iterations at the LT decoder, the check-to-variable message from an output check node, $m_{\mathrm{c} \rightarrow \mathrm{v}}^{(\ell)}$, is converging to the same distribution as the channel message   received at the output bit nodes, $m_0$.  Thus, the  decoded LLR of input bit node $\mathrm{v}$ given in (\ref{decoded_LLR}) can be rewritten  as $\mathcal{L}_{\mathrm{v}} = \sum_{\mathrm{c} \in C_{\mathrm{v}}} m_0$.  Moreover, condition $ \min (\mathcal{L}_\mathrm{v}) \geq \mu_0$ must hold 
to ensure a successful decoding at the LDPC decoder. This alternatively shows that the value of  $\mu_0$  depends on the distribution of the  channel message received at the output bit nodes, which is a symmetric Gaussian distribution with mean, $2 \gamma$,   and variance, $4 \gamma$, where $\gamma$  is the channel SNR.

\subsubsection{Selection of the LDPC code rate}
Previous work concerning Raptor code design appears to have overlooked the choice of LDPC code rate and have selected an arbitrary  LDPC code rate as long as the rate loss is not significant. Mostly Raptor code designers have selected higher rates for the LDPC code such as  0.98 or 0.95 due to  very small rate loss. Cheng~\emph{et~al.}~\cite{chengTcom2009design} showed that a Raptor code with a rate-0.7 LDPC code perform poorly compared to a Raptor code with a rate-0.95 LDPC code, particularly at high SNRs.  
This mainly due to the severe rate loss in the low-rate precoding.  
Generally, at a higher SNR, an LDPC code only performs  error detection rather than error correction. Thus the use of low-rate precoding at high SNRs will result in  severe rate loss. In contrast, at a lower SNR, a low-rate LDPC code can perform both error detection and error correction, thus can help to increase the  realized rate of the LT code component. However, in general the rate gain obtain from low rate precoding is insignificant compared to the rate loss in the entire Raptor code due to low rate precoding. Thus Raptor code designers always encourage to use high-rate LDPC code for precoding.

\section{Results and Discussion}\label{sec-Discussion}

\subsection{Performance comparison of Raptor codes designed using the multi-edge framework and tandem decoding with existing Raptor code results }

In this section, we compare the performance of Raptor codes designed with the  optimization method given in Section~\ref{sec:problem stat for ficed precode} using the multi-edge framework,  with the existing Raptor codes in literature. For a fair comparison, we set same precode settings as the  reference codes  and use tandem decoding scheme as the decoding scheme in the code optimization. 
We consider two reference codes from  literature, where reference code 1~\cite{shirvanimoghaddamTcom2015raptor} was designed  with precode rate of 0.98 for very low SNRs and reference code 2~\cite{etesamiIT2006raptor} was designed with precode rate of 0.98 at SNR 0.5 dB.  We use (4, 200)-regular LDPC code as  the precode  and set maximum number of BP decoding iterations to 1000. We use the maximum average message mean, $\mu_0$, as the switching criterion  from LT code component to LDPC code component with  tandem decoding and set the value of $\mu_0$  to 40 for optimized code 1 and 30 for optimized code 2. In MET-DE we set the number of sample points in the quantization  to 3000 and the quantization interval to 0.01.   

\begin{figure}[!t]
	\centering
	\includegraphics[width=0.6\linewidth]{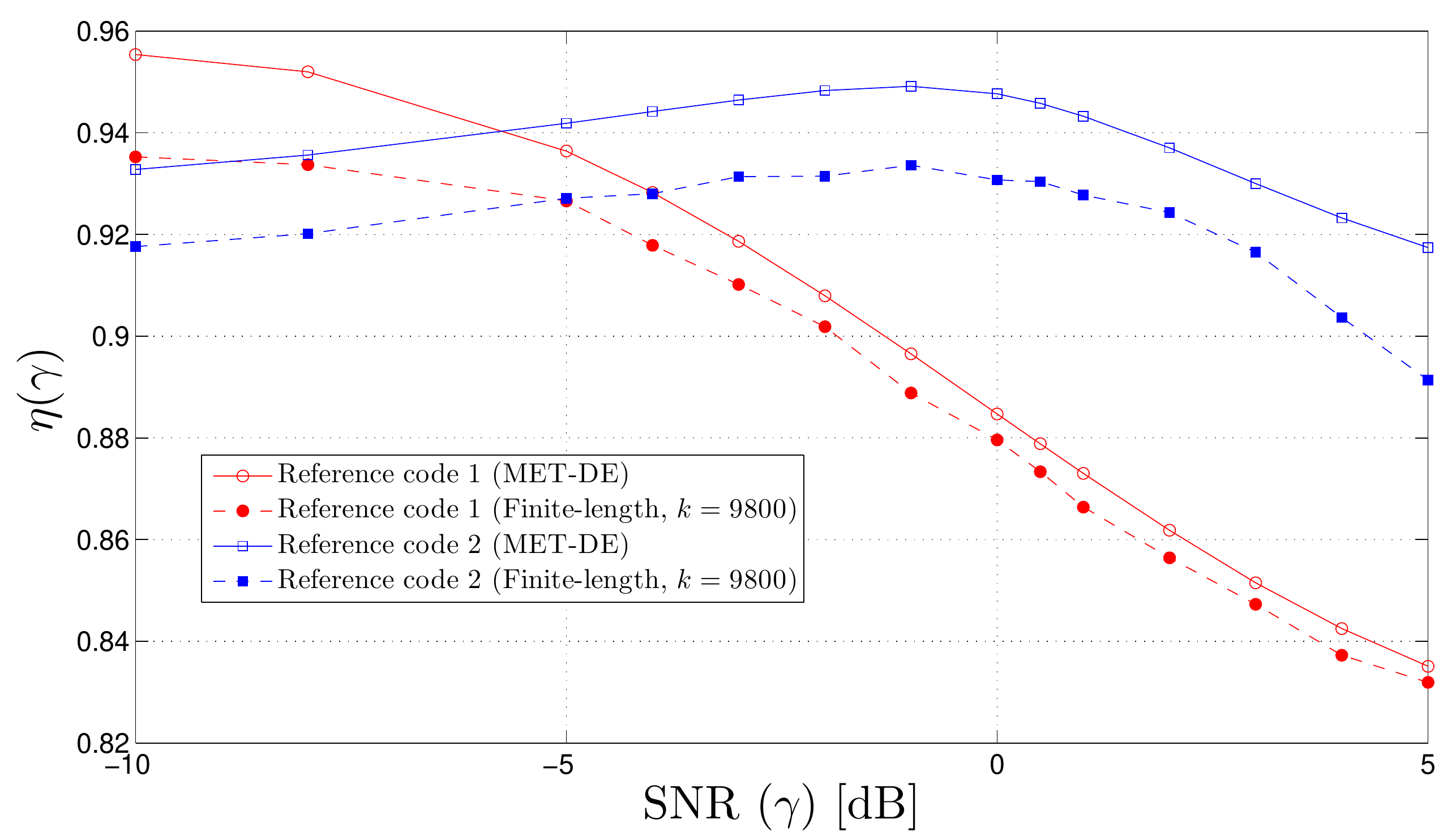}
	\vspace*{-1em}
	\caption{Rate efficiency  results computed  using  MET-DE  and finite-length simulations   for  the existing Raptor codes in literature with fixed LDPC code rate of 0.98, evaluated at  different SNRs. }
	\label{fig:fig5optimizedcodesref}
\end{figure}

In order to verify the correctness of MET-DE for Raptor code analysis, we first compare the rate efficiency results  computed with MET-DE and  finite-length simulations for the existing Raptor codes in literature in Fig.~\ref{fig:fig5optimizedcodesref}.  It is clear from Fig.~\ref{fig:fig5optimizedcodesref} that the rate efficiency results computed with MET-DE  closely follow the  finite-length simulations at all SNRs. We then compare the rate efficiency result of the Raptor codes designed in the multi-edge framework using tandem decoding, with the  existing Raptor codes in literature in Table~\ref{tab:compare_raptor}. It is clear from    Table~\ref{tab:compare_raptor} that  optimized Raptor codes using the multi-edge framework give higher rate efficiencies than  reference codes.
Moreover, optimized codes have   smaller maximum output check node degree compared to reference codes. The advantage of having lower  degrees in a code ensemble is that it helps to reduce the decoding complexity.

\begin{table}[!t]
	\renewcommand{\arraystretch}{1.3}
	\caption{Comparison of optimized Raptor code degree distributions with existing Raptor code degree distributions} \vspace{-1em}
	\label{tab:compare_raptor}
	\centering
	\scriptsize
	\begin{tabular}{|c |c| l| c|c| }
		\hline
		\multicolumn{1}{|c|}{\multirow{2}[-2]{*}{Design }} & 	\multicolumn{1}{c|}{\multirow{2}[2]{*}{Raptor code}} & \multicolumn{1}{c|}{\multirow{2}[2]{*}{Degree distribution}} & \multicolumn{1}{c|}{\multirow{2}[-2]{*}{Avg. output}} & \multicolumn{1}{c|}{\multirow{2}[-2]{*}{$\eta(\gamma_d)$}}\\
		SNR ($\gamma_d$) & & & node degree ($\beta$) & (MET-DE) \\
		\hline
		\hline
		\multicolumn{1}{|c|}{\multirow{4}[2]{*}{-10 dB}} & \multicolumn{1}{c|}{Reference code 1} & $ \Omega(x)  =  0.0174\, x+ 0.3488\, x^2+ 0.2309\, x^3+ 0.0695\, x^4+ 0.0873\, x^5$ & \multicolumn{1}{c|}{\multirow{2}[2]{*}{14.01}} & \multicolumn{1}{c|}{\multirow{2}[2]{*}{\textbf{0.9556}}}\\
		\multicolumn{1}{|c|}{\multirow{4}[2]{*}{}} & \cite[Table II]{shirvanimoghaddamTcom2015raptor} & $ + 0.0002\, x^6 + 0.0805\, x^7+ 0.0004\, x^8+ 0.0191\, x^{11}+ 0.0518\, x^{12}$ &\multicolumn{1}{c|}{\multirow{2}[2]{*}{}} & \multicolumn{1}{c|}{\multirow{2}[2]{*}{}}\\
		\multicolumn{1}{|c|}{\multirow{4}[2]{*}{}} & & $+ 0.0123\, x^{23}+ 0.031\, x^{24} + 0.022\, x^{59}+ 0.002\, x^{60}+ 0.02683\, x^{300}$ &\multicolumn{1}{c|}{\multirow{2}[2]{*}{}} & \multicolumn{1}{c|}{\multirow{2}[2]{*}{}}\\
		\cline{2-5}
		\multicolumn{1}{|c|}{\multirow{3}[2]{*}{}} & \multicolumn{1}{c|}{\multirow{2}[2]{*}{Optimized code 1}} & $ \Omega(x)  = 0.0261\, x+ 0.3526\, x^2+ 0.3195\, x^3+ 0.0946\, x^5+ 0.0076\, x^6 $ & \multicolumn{1}{c|}{\multirow{2}[2]{*}{6.68}}& \multicolumn{1}{c|}{\multirow{2}[2]{*}{\textbf{0.9606}}}\\
		\multicolumn{1}{|c|}{\multirow{2}[2]{*}{}} &  & $+ 0.1508\, x^{12}+ 0.0055\, x^{16}+ 0.002\, x^{19}+ 0.008\, x^{23}+ 0.0091\, x^{61}$  &\multicolumn{1}{c|}{\multirow{2}[2]{*}{}}&  \multicolumn{1}{c|}{\multirow{2}[2]{*}{}}\\
		\multicolumn{1}{|c|}{\multirow{2}[2]{*}{}} &  & $+ 0.023\, x^{63}+ 0.00123\, x^{290}$  &\multicolumn{1}{c|}{\multirow{2}[2]{*}{}}&  \multicolumn{1}{c|}{\multirow{2}[2]{*}{}}\\
		
		\hline
		\multicolumn{1}{|c|}{\multirow{4}[2]{*}{0.5 dB}} & \multicolumn{1}{c|}{Reference code 2} & $ \Omega(x)  = 0.0006\, x+ 0.492\, {x}^2+ 0.0339\, {x}^3+ 0.2403\, {x}^4+ 0.006\, {x}^5   $ & \multicolumn{1}{c|}{\multirow{2}[2]{*}{11.16}}& \multicolumn{1}{c|}{\multirow{2}[2]{*}{\textbf{0.9458}}}\\
		\multicolumn{1}{|c|}{\multirow{4}[2]{*}{}} & \cite[Page 2044]{etesamiIT2006raptor} & $+ 0.095\, {x}^8+ 0.049\, {x}^{14}+ 0.018\, {x}^{30}+ 0.0356\, {x}^{33}+ 0.0296\, {x}^{200}  $ & \multicolumn{1}{c|}{\multirow{2}[2]{*}{}}& \multicolumn{1}{c|}{\multirow{2}[2]{*}{}}\\
		\cline{2-5}
		\multicolumn{1}{|c|}{\multirow{4}[2]{*}{}} & \multicolumn{1}{c|}{\multirow{2}[2]{*}{Optimized code 2}} & $ \Omega(x)  = 0.0082\, x+ 0.5019\, x^2+ 0.043\, x^3+ 0.2365\, x^4+ 0.0067\, x^5$ & \multicolumn{1}{c|}{\multirow{2}[2]{*}{11.47}}&\multicolumn{1}{c|}{\multirow{2}[2]{*}{\textbf{0.9524}}}\\
		\multicolumn{1}{|c|}{\multirow{2}[2]{*}{}} &  & $+ 0.0911\, x^8 + 0.0398\, x^{14}+ 0.0108\, x^{30}+ 0.0273\, x^{33}+ 0.0347\, x^{197}$ & \multicolumn{1}{c|}{\multirow{2}[2]{*}{}} & \multicolumn{1}{c|}{\multirow{2}[2]{*}{}}\\
		\hline								
	\end{tabular}
	\vspace*{-2em}
\end{table}

\normalsize
\vspace*{-1em}

\subsection{Tandem decoding vs. joint decoding} \label{sec:fixed precode}

\begin{figure}[!t]
	\centering
	\includegraphics[width=0.6\linewidth]{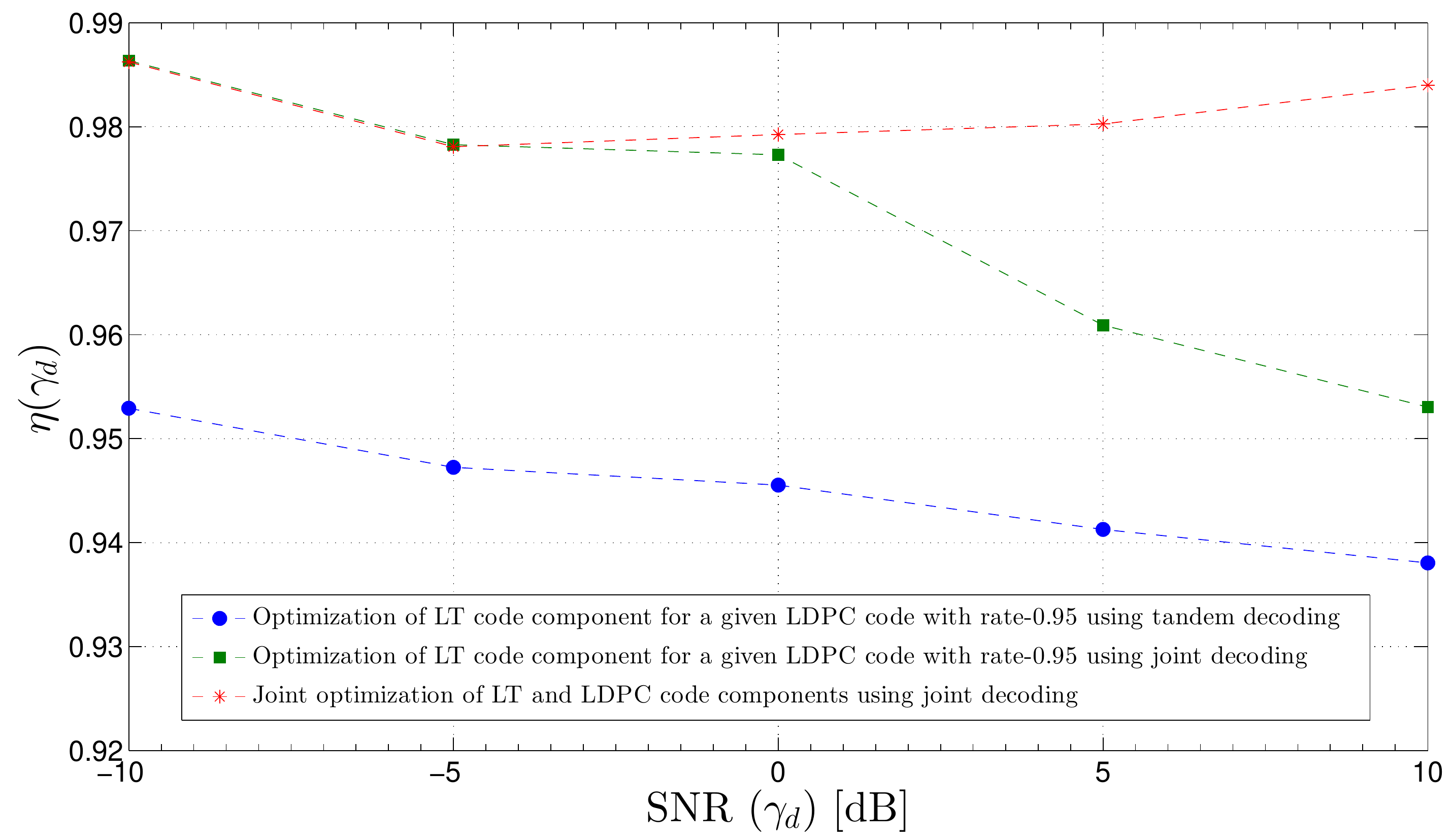}
	\vspace*{-1em}
	\caption{Rate efficiency results computed using MET-DE for the  Raptor codes designed using the multi-edge framework, evaluated at the designed SNR. }
	\label{fig:optimizedcodefinal}
\end{figure}

In this section, we consider the design of Raptor codes for a given precode rate using tandem and joint decoding schemes. We formulate the optimization problem as per Section~\ref{sec:problem stat for ficed precode} and set the precode to a (3,60)-regular LDPC code of rate 0.95, and maximum check node degree, $j_{c_{\max}}$, to  50.  In Fig.~\ref{fig:optimizedcodefinal} we show the rate efficiency results of  the Raptor code degree distributions designed  using joint and tandem decoding schemes  for  different SNRs. Note that  each point in the figure corresponds to a code designed specifically for that SNR. It can be observed  that  Raptor codes designed using joint decoding  always outperform  Raptor codes designed using tandem decoding in terms of rate efficiency. However, as the designed SNR increases, the rate efficiency gap between  Raptor codes designed with joint decoding and tandem decoding is  reduced.  Furthermore with  joint decoding, the BP decoding algorithm converges to a zero BER faster than with  tandem decoding, which improves the rate efficiency. Additional advantages of  joint decoding is that  we no longer need to consider the switching point between LT code component to LDPC code component, and  we can use the LDPC code parity-check to halt decoding early once a valid codeword   is found.

\begin{figure}[!t]
	\centering
	\includegraphics[width=0.6\linewidth]{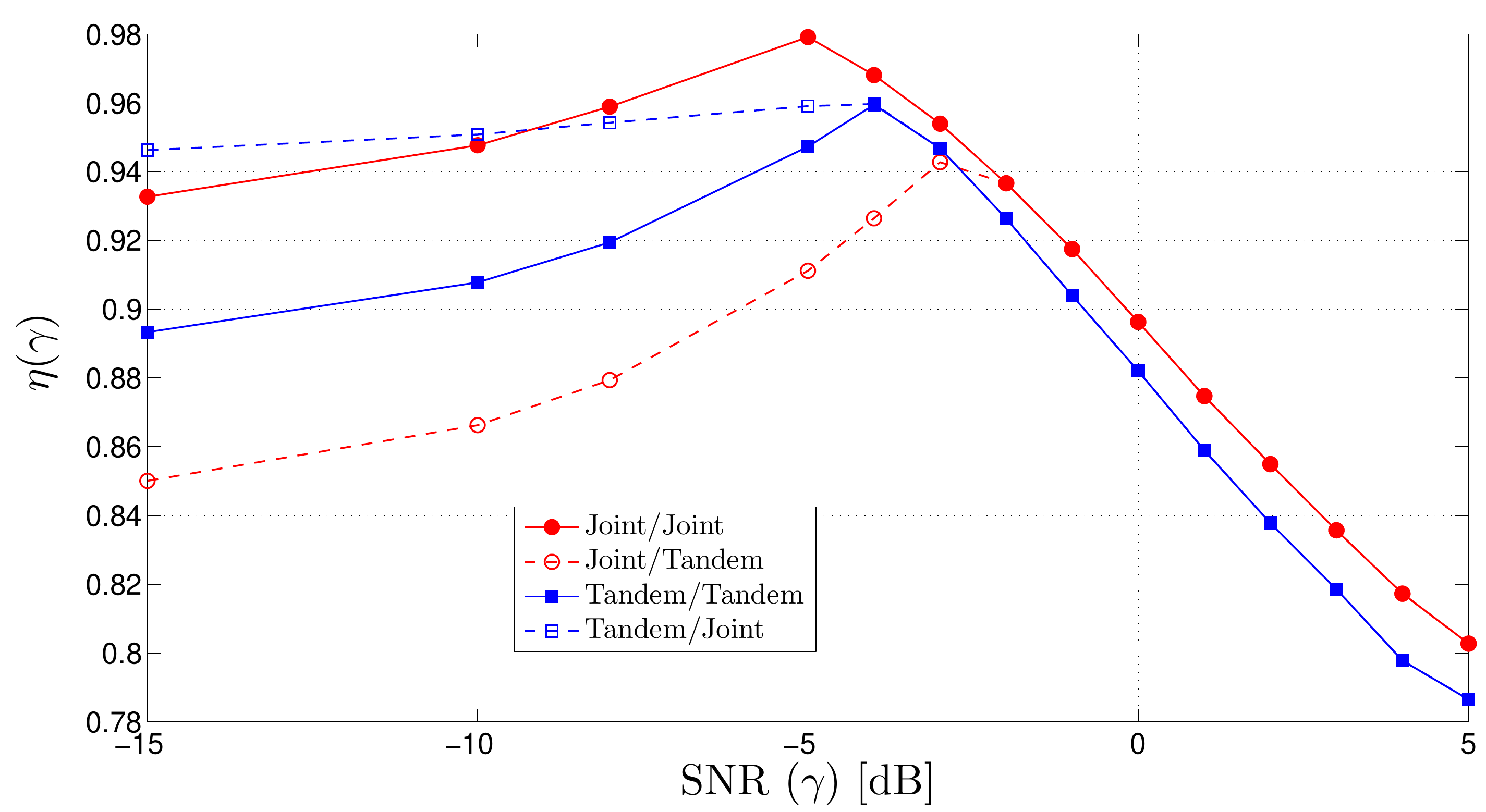}
	\vspace*{-1em}
	\caption{Rate efficiency  results computed using MET-DE for the  Raptor codes designed at -5dB with fixed LDPC code rate of 0.95, designed/evaluated using joint decoding  and tandem decoding.}
	\label{fig:optimized_code_vs_SNR} 
\end{figure}

In Table~\ref{tab:compare_tandem_joint} we compare the rate efficiency results computed using MET-DE for the optimized Raptor codes shown in Fig.~\ref{fig:optimizedcodefinal} with fixed LDPC code rate of 0.95, designed/evaluated using joint decoding  and tandem decoding at the designed SNR. 
It is clear from    Table~\ref{tab:compare_tandem_joint} that  Raptor codes designed and/or evaluated  using joint decoding always show the best rate efficiency  performance. 
Moreover in Fig.~\ref{fig:optimized_code_vs_SNR},  we evaluate the rate efficiency performance of optimized degree distributions designed at -5 dB (which are shown in Fig.~\ref{fig:optimizedcodefinal} with fixed LDPC code rate of 0.95)  for SNRs above and below the designed SNR using tandem decoding and joint decoding. It can be observed from Fig.~\ref{fig:optimized_code_vs_SNR} that Raptor codes evaluated using joint decoding  always show the best rate efficiency performance regardless of whether the code was optimized with joint or tandem decoding. This seems to be significant when the evaluated SNR is below  the designed SNR. The reason is that the use of the information generated from both LT and LDPC code components in parallel helps to minimize the residual error; thus improving the rate efficiency performance.

\begin{table}[!t]
	\renewcommand{\arraystretch}{1.3}
	\caption{Rate efficiency  results computed using MET-DE for the optimized Raptor codes shown in Fig.~\ref{fig:optimizedcodefinal} with fixed LDPC code rate of 0.95, designed/evaluated using joint decoding  and tandem decoding at the designed SNR.} \vspace{-1em}
	\label{tab:compare_tandem_joint}
	\centering
	\scriptsize
	\begin{tabular}{|c |c| c| c| c|}
		\hline	
		Designed & \multicolumn{4}{c|}{$\eta(\gamma_d)$ (MET-DE)}	\\
		\cline{2-5}		
		SNR ($\gamma_d$) & Joint/Joint & Joint/Tandem & Tandem/Tandem & Tandem/Joint \\
		\hline
		-10 dB  & 0.9864 & 0.9053 & 0.9529 & 0.9571 \\
		-5  dB  & 0.9783 & 0.9112 & 0.9472 & 0.9590 \\
		0  dB   & 0.9773 & 0.9173 & 0.9455 & 0.9633 \\
		5  dB   & 0.9609 & 0.9159 & 0.9413 & 0.9541 \\
		10  dB  & 0.9530 & 0.9174 & 0.9380 & 0.9472 \\	
		\hline						
	\end{tabular}
	\vspace*{-2em}
\end{table}

\vspace*{-1em}

\subsection{The Raptor codes design as MET-LDPC codes using joint decoding }

In this section, we consider the design of Raptor codes as  MET-LDPC codes, where we do not fix the rate and the structure of the precode in advance. We formulate the optimization problem as per Section~\ref{sec:optimzation sta MET}. The main advantage of this method is that, it allows us to select the most suitable precode setting depending on the design SNR. 

The degree distributions of Raptor codes designed with the multi-edge framework using joint decoding scheme for different SNRs are given in Table~\ref{Table : joint decoding}. We present the degree distributions using MET notations as we design them as MET-LDPC codes. One can easily compute the relevant  $\Omega(x)$ and the degree distributions of LDPC  code using (\ref{omega_x}) and (\ref{LDPC_dd}). 
In Fig.~\ref{fig:optimizedcodefinal} we show the rate efficiency results for the optimized Raptor codes  given in Table~\ref{Table : joint decoding}  to compare the rate efficiency results obtained with fixed precode setting. 
It is clear from  results shown in this section that 
the Raptor codes designed using the multi-edge framework achieve realized rates close to the channel capacity for any channel SNR (Indeed codes can be designed at any SNR regardless of whether joint or tandem decoding is considered). Therefore, the proposed method using the multi-edge framework succeeds to overcome the problem in existing   linear programming based on mean-LLR-EXIT chart method~\cite{etesamiIT2006raptor,chengTcom2009design}, where it is only capable of giving the capacity achieving codes between two SNR bounds $\text{SNR}^*_{\text{low}}$ and $\text{SNR}^*_{\text{high}}$ given in~\cite{chengTcom2009design}. 

\begin{table}[!t]
	\renewcommand{\arraystretch}{1.3}
	\caption{Optimized degree distributions of Raptor codes as MET-LDPC codes for different designed SNRs using joint decoding} \vspace{-1em}
	\label{Table : joint decoding}
	\centering
	\scriptsize
	\begin{tabular}{|c |l| c| c| }
		\hline
		\multicolumn{1}{|c|}{\multirow{2}[-2]{*}{Designed }} & 	\multicolumn{1}{c|}{\multirow{2}[2]{*}{Raptor code as a MET-LDPC code}} & \multicolumn{1}{c|}{\multirow{2}[2]{*}{$\mathcal{R}_{\text{LDPC}}$}} & \multicolumn{1}{c|}{\multirow{2}[-2]{*}{$\eta(\gamma_d)$}}\\
		SNR ($\gamma_d$) & & & (MET-DE)\\ 
		\hline
		\hline
		\multicolumn{1}{|c|}{\multirow{2}[2]{*}{-10 dB}} & $L(\boldsymbol{r},\boldsymbol{x}) =   0.0445\, {x_{1}}^3\, {x_{2}}^{61}+ 0.0268\, {x_{1}}^3\, {x_{2}}^{62}+x_{3} $ & \multicolumn{1}{c|}{\multirow{2}[2]{*}{0.9500}} & \multicolumn{1}{c|}{\multirow{2}[2]{*}{\textbf{0.9862}}}\\
		\multicolumn{1}{|c|}{\multirow{2}[-3]{*}{}} & $R(\boldsymbol{x}) =  0.0036\, {x_{1}}^{60}+  0.0117\, x_{2}\, x_{3}+ 0.3616\, {x_{2}}^2\, x_{3}+ 0.3005\, {x_{2}}^3\, x_{3}+ 0.0958\, {x_{2}}^4\, x_{3} $ & \multicolumn{1}{c|}{\multirow{2}[2]{*}{}} & \multicolumn{1}{c|}{\multirow{2}[2]{*}{}}\\
		\multicolumn{1}{|c|}{\multirow{2}[-3]{*}{}} & $+ 0.0952\, {x_{2}}^7\, x_{3}+ 0.0344\, {x_{2}}^9\, x_{3}+ 0.046\, {x_{2}}^{12}\, x_{3}+ 0.0389\, {x_{2}}^{12}\, x_{3}+ 0.0149\, {x_{2}}^{21}\, x_{3}$ & \multicolumn{1}{c|}{\multirow{2}[2]{*}{}} & \multicolumn{1}{c|}{\multirow{2}[2]{*}{}}\\	
		\multicolumn{1}{|c|}{\multirow{2}[-3]{*}{}} & $+ 0.001\, {x_{2}}^{50}\, x_{3}$ & \multicolumn{1}{c|}{\multirow{2}[2]{*}{}} & \multicolumn{1}{c|}{\multirow{2}[2]{*}{}}\\					
		\hline
		
		\multicolumn{1}{|c|}{\multirow{2}[2]{*}{-5 dB}} & $L(\boldsymbol{r},\boldsymbol{x}) =  0.054\, {x_{1}}^3\, {x_{2}}^{19}+ 0.1506\, {x_{1}}^3\, {x_{2}}^{20}+ x_{3}$ & \multicolumn{1}{c|}{\multirow{2}[2]{*}{0.9450}} & \multicolumn{1}{c|}{\multirow{2}[2]{*}{\textbf{0.9781	}}}\\
		\multicolumn{1}{|c|}{\multirow{2}[2]{*}{}} & $R(\boldsymbol{x}) =  0.006139\, {x_{1}}^{55}+ 0.0051\, {x_{1}}^{54}+ 0.0436\, x_{2}\, x_{3}+ 0.3796\, {x_{2}}^3\, x_{3}+ 0.2738\, {x_{2}}^2\, x_{3}$& \multicolumn{1}{c|}{\multirow{2}[2]{*}{}} & \multicolumn{1}{c|}{\multirow{2}[2]{*}{}}\\
		\multicolumn{1}{|c|}{\multirow{2}[2]{*}{}} & $+ 0.0048\, {x_{2}}^4\, x_{3}+ 0.0386\, {x_{2}}^2\, x_{3}+ 0.022\, {x_{2}}^3\, x_{3}+ 0.1211\, {x_{2}}^{10}\, x_{3}+ 0.1057\, {x_{2}}^6\, x_{3}$ & \multicolumn{1}{c|}{\multirow{2}[2]{*}{}} & \multicolumn{1}{c|}{\multirow{2}[2]{*}{}}\\
		\multicolumn{1}{|c|}{\multirow{2}[2]{*}{}} & $+ 0.0104\, {x_{2}}^{27}\, x_{3}+ 0.0004\, {x_{2}}^{50}\, x_{3}$ & \multicolumn{1}{c|}{\multirow{2}[2]{*}{}} & \multicolumn{1}{c|}{\multirow{2}[2]{*}{}}\\
		\hline
		
		\multicolumn{1}{|c|}{\multirow{2}[2]{*}{0 dB}} & $L(\boldsymbol{r},\boldsymbol{x}) =  0.0047\, {x_{1}}^3\, {x_{2}}^6+0.4904\, {x_{1}}^3\, {x_{2}}^7+  x_{3}$ & \multicolumn{1}{c|}{\multirow{2}[2]{*}{0.9610}} & \multicolumn{1}{c|}{\multirow{2}[2]{*}{\textbf{0.97923	}}}\\
		\multicolumn{1}{|c|}{\multirow{2}[2]{*}{}} & $R(\boldsymbol{x}) = 0.0178\, {x_{1}}^{77}+ 0.0014\, {x_{1}}^{76}+ 0.0318\, x_{2}\, x_{3}+ 0.3955\, {x_{2}}^2\, x_{3}+ 0.2926\, {x_{2}}^3\, x_{3}$ & \multicolumn{1}{c|}{\multirow{2}[2]{*}{}} & \multicolumn{1}{c|}{\multirow{2}[2]{*}{}}\\
		\multicolumn{1}{|c|}{\multirow{2}[2]{*}{}} & $+ 0.0738\, {x_{2}}^3\, x_{3}+ 0.0711\, {x_{2}}^3\, x_{3}+ 0.0575\, {x_{2}}^9\, x_{3}+ 0.0528\, {x_{2}}^9\, x_{3}+ 0.0236\, {x_{2}}^{12}\, x_{3}$ & \multicolumn{1}{c|}{\multirow{2}[2]{*}{}} & \multicolumn{1}{c|}{\multirow{2}[2]{*}{}}\\
		\multicolumn{1}{|c|}{\multirow{2}[2]{*}{}} & $+ 0.0006\, {x_{2}}^{25}\, x_{3}+ 0.0007\, {x_{2}}^{50}\, x_{3}$ & \multicolumn{1}{c|}{\multirow{2}[2]{*}{}} & \multicolumn{1}{c|}{\multirow{2}[2]{*}{}}\\
		\hline
		
		\multicolumn{1}{|c|}{\multirow{2}[2]{*}{5 dB}} & $L(\boldsymbol{r},\boldsymbol{x}) = 0.1122\, {x_{1}}^2\, {x_{2}}^3+ 0.1295\, {x_{1}}^3\, {x_{2}}^3+ 0.6092\, {x_{1}}^3\, {x_{2}}^4+ x_{3} $ & \multicolumn{1}{c|}{\multirow{2}[2]{*}{0.9898}} & \multicolumn{1}{c|}{\multirow{2}[2]{*}{\textbf{0.9802}}}\\
		\multicolumn{1}{|c|}{\multirow{2}[-3]{*}{}} & $R(\boldsymbol{x}) =  0.0014\, {x_{1}}^{297}+ 0.0016\, {x_{1}}^{297}+ 0.0025\, {x_{1}}^{300}+ 0.0026\, {x_{1}}^{299}+ 0.0005\, {x_{1}}^{44} $ & \multicolumn{1}{c|}{\multirow{2}[2]{*}{}} & \multicolumn{1}{c|}{\multirow{2}[2]{*}{}}\\
		\multicolumn{1}{|c|}{\multirow{2}[-3]{*}{}} & $+ 0.0915\, x_{2}\, x_{3} + 0.3848\, {x_{2}}^2\, x_{3}+ 0.4328\, {x_{2}}^3\, x_{3}+ 0.026\, {x_{2}}^4\, x_{3}+ 0.0109\, {x_{2}}^7\, x_{3}		$ & \multicolumn{1}{c|}{\multirow{2}[2]{*}{}} & \multicolumn{1}{c|}{\multirow{2}[2]{*}{}}\\
		\multicolumn{1}{|c|}{\multirow{2}[-3]{*}{}} & $+ 0.0258\, {x_{2}}^{10}\, x_{3}+ 0.0282\, {x_{2}}^{20}\, x_{3} $ & \multicolumn{1}{c|}{\multirow{2}[2]{*}{}} & \multicolumn{1}{c|}{\multirow{2}[2]{*}{}}\\
		\hline
		
		\multicolumn{1}{|c|}{\multirow{2}[2]{*}{10 dB}} & $L(\boldsymbol{r},\boldsymbol{x}) =  0.3301\, x_{1}\, {x_{2}}^2+ 0.297\, x_{1}\, {x_{2}}^3+ 0.3604\, {x_{1}}^2\, {x_{2}}^3+ x_{3} $ & \multicolumn{1}{c|}{\multirow{2}[2]{*}{0.9932 }} & \multicolumn{1}{c|}{\multirow{2}[2]{*}{\textbf{0.9840}}}\\
		\multicolumn{1}{|c|}{\multirow{2}[-3]{*}{}} & $R(\boldsymbol{x}) = 0.0024\, {x_{1}}^{242}+ 0.0021\, {x_{1}}^{80}+ 0.0015\, {x_{1}}^{300}+ 0.0001\, {x_{1}}^{232}+ 0.0006\, {x_{1}}^{210}$ & \multicolumn{1}{c|}{\multirow{2}[2]{*}{}} & \multicolumn{1}{c|}{\multirow{2}[2]{*}{}}\\
		\multicolumn{1}{|c|}{\multirow{2}[-3]{*}{}} & $+~ 0.0543\, x_{2}\, x_{3}+ 0.6645\, {x_{2}}^2\, x_{3}+ 0.1362\, {x_{2}}^3\, x_{3}+ 0.0943\, {x_{2}}^4\, x_{3}+ 0.0367\, {x_{2}}^5\, x_{3}$ & \multicolumn{1}{c|}{\multirow{2}[2]{*}{}} & \multicolumn{1}{c|}{\multirow{2}[2]{*}{}}\\
		\multicolumn{1}{|c|}{\multirow{2}[-3]{*}{}} & $+ 0.014\, {x_{2}}^{20}\, x_{3}  $ & \multicolumn{1}{c|}{\multirow{2}[2]{*}{}} & \multicolumn{1}{c|}{\multirow{2}[2]{*}{}}\\
		\hline				
	\end{tabular}
\end{table}
\normalsize

\vspace*{-1em}

\subsection{Incremental decoding strategy for Raptor codes designed using tandem decoding scheme}  

This section considers the incremental decoding strategy for Raptor codes. 
For simplicity, we assume that all decoding attempts are uniformly spaced, i,e.,  the number of extra output bits ($\delta m$) received at each decoding attempt is the same. This is true for  both message-reset decoding and incremental decoding. We also assume that the LT decoding process will perform a predetermined number of BP decoding iterations ($T$).  
The decoding complexity of  the message-reset decoding and   incremental decoding   is given in Table~\ref{Table : FP values}. This gives the number of floating point operations per decoding attempt, which is  also the number of convolutions involved in the MET-DE.  It is clear from  Table~\ref{Table : FP values} that the incremental  decoding strategy  has a lower decoding complexity compared to the message-reset decoding, because the Tanner graph corresponding to  incremental decoding has a smaller number of edges compared to the one with message-reset decoding. Note that if we set $x$ to one, then the decoding complexity of the check node side is the same for both incremental decoding and message-reset decoding.

\begin{table}[!t]
	\renewcommand{\arraystretch}{1.5}
	\caption{Floating point operations per decoding attempt for the LT code component of the Raptor code} \vspace{-1em}
	\label{Table : FP values}
	\centering	
	\scriptsize
	\begin{center}
		\begin{tabular}{|c|c|c|}
			\hline 
			Decoding strategy & Variable nodes  & Check nodes  \\ 
			\hline 
			Message-reset decoding\footnotemark[1] & $(n i_v + m + \delta m)T$  & $(m j_{c} +  \delta m d_{c})T $ \\ 
			\hline 
			Incremental decoding\footnotemark[2]& $(n \bar{i}_v + xm + \delta m)T$  & $(x m \bar{j}_{c} +  \delta m \bar{d}_{c})T $  \\ 
			\hline 
		\end{tabular} 
	\end{center}
	\begin{tablenotes}
		\small
		\item {\footnotemark[1] \scriptsize $i_v$, $j_{c}$ and $d_{c}$ respectively denote the average degree of input bit nodes, average degree of output check nodes received during previous decoding attempts and  the average degree of output check nodes received at the current decoding attempt for  message-reset decoding.  }
		\item {\footnotemark[2] \scriptsize $\bar{i}_v$, $\bar{j}_{c}$ and $\bar{d}_{c}$ respectively denote the average degree of input bit nodes, average degree of the fraction of output check nodes selected from previous decoding attempts and  the average degree of output check nodes received at the current decoding attempt for  incremental decoding.  Note that $i_v > \bar{i}_v$ and $j_{c} > \bar{j}_{c}$ for $x<1$.}
	\end{tablenotes}
	\vspace*{-2em}
\end{table}

\begin{figure}[!t]
\centering
\includegraphics[width=0.6\linewidth]{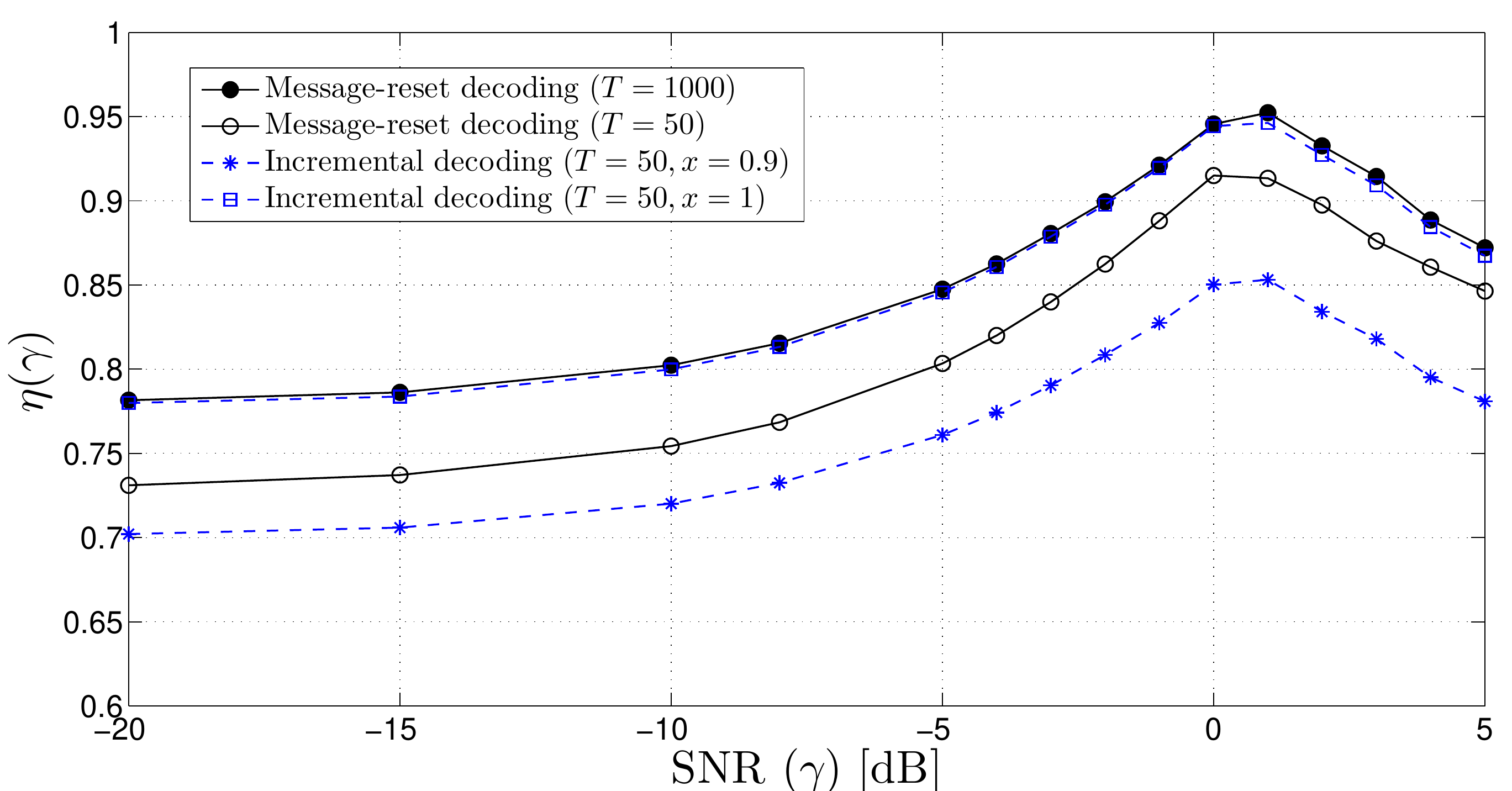}
\caption{Rate efficiency performance of the Raptor code designed at 0 dB with tandem decoding scheme for fixed precode settings in Section~\ref{sec:fixed precode}, evaluated with incremental decoding strategy.}
\label{fig:fig5on-the-flyopt}
\end{figure}

\normalsize
We then evaluate the  performance of these two decoding strategies  using the rate efficiency results.  We consider the Raptor code designed at 0 dB with tandem decoding scheme for fixed precode settings in Section~\ref{sec:fixed precode}, which has LT degree distribution of 
\begin{multline}
\label{omega}
\Omega(x) = 0.0135x +   0.4920x^2 +  0.2836x^3  +  0.0538x^5  +  0.0499x^8  +  0.0779x^9  +  \\ 0.0287 x^{19} +  0.0006 x^{50}
\end{multline}
In Fig.~\ref{fig:fig5on-the-flyopt} we show the rate efficiency results for (\ref{omega}) decoded with the incremental decoding strategy.  For a fair comparison, we also show the  rate efficiency results for (\ref{omega}) decoded with  message-reset decoding using the same   number of BP decoding iterations ($T$). 
It is clear from Fig.~\ref{fig:fig5on-the-flyopt} that  incremental decoding ($T=50$ and $x=1$) outperforms   message-reset decoding ($T=50$) with the same decoding complexity. Moreover, we can obtain rate efficiency results  close to the optimal result (i.e., message-reset decoding with $T=1000$,  where we have chosen $T=1000$ as increasing  $T$ beyond 1000 does not noticeably improve $\eta(\gamma)$)  by having incremental decoding of $T=50,x=1$, which offers a significant reduction of the decoding complexity.  
This is mainly due to the  fact that  recalculation of the same information is avoided with the incremental decoding.
Therefore, we can suggest that  incremental decoding closely approximate the optimal decoding strategy  for the Raptor code design, where accurate and efficient rate calculation is particularly valuable.

\vspace*{-1em}

\section{Conclusion}\label{sec-conclusion}
In this paper, we proposed a new representation of Raptor codes as a multi-edge type low-density parity-check (MET-LDPC) code. 
This MET representation  enable us to perform a comprehensive analysis of asymptotic performance of the Raptor code using  MET density evolution (MET-DE).
The advantage of using MET-DE over conventional Gaussian approximation-based approaches, such as  mean-LLR-EXIT chart,  is that  MET-DE does not incorporate any Gaussian approximation, thus enables to find the optimal Raptor code  for given channel condition. In addition, unlike finite-length analysis MET-DE gives  the average performance of Raptor code ensemble. 
We  considered two decoding schemes based on the belief propagation  decoding, namely  tandem decoding and joint decoding in the multi-edge framework, and analyzed the stability conditions using MET-DE.  
We then formulated the optimization problem to design  Raptor codes using the multi-edge framework. 
Finally, we considered performance--complexity trade--offs of Raptor codes using the multi-edge framework. The density evolution analysis showed that  Raptor codes designed with joint decoding always outperform the ones designed with tandem decoding. In several examples, we demonstrated that Raptor codes designed using the multi-edge framework   outperform previously reported Raptor codes.

\appendices
\vspace*{-1em}

\section{} \label{appendix_A}

\begin{lemma}[Degree-one variable nodes~\cite{RichardsonW2002multi}]\label{Degree-one}
	Any check node that receives information from a degree-one variable node  outputs a check-to-variable message with the distribution $\Delta$ in all the edges except the edge which connects to the degree-one variable node if that check node receives messages from its connected variable nodes with a distribution of infinitely large mean. 
	The distribution $\Delta$ is converging to  the same distribution as the channel message received to its connected degree-one variable node as the number of BP decoding iterations goes to infinity. 
\end{lemma}
\begin{IEEEproof}
	Consider a set of independent random variables $x_1 \dots x_n$, and let $\mathcal{L}(x_i)$ denote the corresponding LLR value of random variable $x_i$. Then the modulo-2-sum  of  LLRs of $x_1 \dots x_n$ is 
	$\mathcal{L}(x_1)\oplus\mathcal{L}(x_2) \oplus \dots \oplus  \mathcal{L}(x_n) = \mathcal{L}(x_1 \oplus x_2 \oplus \dots \oplus x_n)$
	and $\mathcal{L}(x_1 \oplus x_2 \oplus \dots \oplus x_n) \stackrel{p}{\rightarrow} \mathcal{L}(x_1)$ as $\min_{2 \leq i \leq n} \mathbb{E}[\mathcal{L}(x_i)] \rightarrow \infty$. 	
\end{IEEEproof}

\begin{remark}[Stability of MET-LDPC codes with degree-one variable nodes~{\protect\cite[pages 396-397]{richardsonBOOK2008modern}}]\label{stability}
	Consider a  MET-LDPC code having no degree-one variable nodes together with its serial concatenation with LDGM code. We refer the sub-Tanner graph having no degree-one variable nodes  as the \emph{Core LDPC graph}. Let $E_1$ denote the set of edge-types in the core LDPC graph, $E_2$ denote the set of edge-types attached exclusively to degree-one variable nodes and $E_{1,2}$ denote the set of edge-types which connect variable nodes in the core LDPC graph to the check nodes associated to the LDGM code. Then if the perfectly decodable fixed point of the core LDPC graph is stable with the PDFs carried on edge-type $E_{1,2}$ in check-to-variable direction, then the fixed point associated to the full graph is stable.  
\end{remark}

\vspace*{-1em}
	
\section{Proof of Theorem~\ref{Stability_Raptor}} \label{appendix_B}
 
Let $a_i^{(\ell)}$ and $b_i^{(\ell)}$ respectively denote the PDFs of  variable-to-check message and check-to-variable message along 
$\Pi_i$  at the $\ell$th BP decoding iteration. Let  $x_i^{(\ell)}$ and $y_i^{(\ell)}$ respectively denote
the Bhattacharyya constants~{\protect\cite[pages 202]{richardsonBOOK2008modern}} associated with $a_i^{(\ell)}$ and $b_i^{(\ell)}$.
Note that $\Pi_1$ represents the core LDPC graph. 
Using equations~(\ref{eq:VN to CN PDF MET}) and~(\ref{eq:CN to VN PDF MET}) we can write  DE equations related to variable nodes and check nodes  of the core LDPC graph as follows: 
\begin{align}
	\label{variable_node_update_core}
	a_1^{(\ell+1)} &=  \sum_{i\geq2} \sum_{j\geq 1} \lambda_{[i~j]}\left[b_1^{(\ell)}\right]^{\otimes (i-1)} 		\otimes\left[b_2^{(\ell)}\right]^{\otimes (j) }  = \lambda_1(b_1^{(\ell)},b_2^{(\ell)}),  \\
	\label{check_node_update_core}
	b_1^{(\ell)} &=  \sum_{i\geq 0} \rho_i\left[a_1^{(\ell)}\right]^{\boxtimes (i-1)} 	= \rho_1(a_1^{(\ell)}),
\end{align}
where  $\lambda_1( x_1, x_2)$ and 	$\rho_1(x_1)$ are given in~(\ref{core_VN}) and~(\ref{core_CN}). Then applying Lemma~4.63 given in~{\protect\cite[pages 202]{richardsonBOOK2008modern}} to (\ref{variable_node_update_core}) and (\ref{check_node_update_core}) give   
\begin{align}
	x_1^{(\ell+1)} &= \sum_{i\geq2} \sum_{j\geq 1} \lambda_{[i~j]}~(y_1^{(\ell)})^{i-1}~(y_2^{(\ell)})^{j} \label{B_1}, \\
	y_1^{(\ell)} &\leq 1-\sum_{i\geq 0} \rho_i (1-x_1^{(\ell)})^{i-1}. \label{B_2}
\end{align}
This finally gives us the update rule for 	$x_1^{(\ell+1)}$ as  
\begin{align}
	x_1^{(\ell+1)} &\leq \sum_{i\geq2} \sum_{j\geq 1} \lambda_{[i~j]}~\left(1-\sum_k \rho_k (1-x_1^{(\ell)})^{k-1}\right)^{i-1}~\left(y_2^{(\ell)}\right)^{j}. \label{B_3}
\end{align}
Furthermore, to ensure that the BER decreases throughout the BP decoding iterations condition, $x_1^{(\ell+1)}< x_1^{(\ell)}$, must hold. Thus for successful decoding under DE, we need to guarantee that 
\begin{align}
	x_1^{(\ell)} &> \sum_{i\geq2} \sum_{j\geq 1} \lambda_{[i~j]}~\left(1-\sum_k \rho_k (1-x_1^{(\ell)})^{k-1}\right)^{i-1}~\left(y_2^{(\ell)}\right)^{j}. \label{B_4}
\end{align}
For the simplicity we  rewire (\ref{B_4}) as   
\begin{align}	
	x_1^{(\ell)} &> f(X_1,X_2), \label{B_5}
\end{align}
where $f(X_1,X_2) = \sum_{i\geq2} \sum_{j\geq 1} \lambda_{[i~j]}~X_1^{i-1}~X_2^j$, and $X_1$ and $X_2$, respectively denote the  $1- \sum_k \rho_k (1-x_1^{(\ell)})^{k-1}  $ and $y_2^{(\ell)}$.
 We  assume that $X_2 = y_2^{(\ell)}= \mathcal{B}(b_2^{(\ell)})$ is independent from $x_1^{(\ell)}$ and $\lim_{\ell\rightarrow \infty}b_2^{(\ell)}$ converges to the channel LLR ($a_0$). Thus $\lim_{\ell\rightarrow \infty}y_2^{(\ell)} = x_0$, where $x_0^{(\ell)}$ is the Bhattacharyya constant associated with $a_0$. For the decoding to be successful, inequality (\ref{B_5}) needs to be valid around  $x_1^{(\ell)}=0$. Thus taking the derivative of  (\ref{B_5}) with respect to $x_1^{(\ell)}=0$ gives us 
\begin{align*}
	\frac{d}{dx_1^{(\ell)}} f(X_1,X_2)\big|_{x_1^{(\ell)}=0} &< 1,
\end{align*}
where \vspace*{-1em}
\begin{align*}
	\frac{d}{dx_1^{(\ell)}} f(X_1,X_2)\big|_{x_1^{(\ell)}=0} &= \lambda_1'(X_1,X_2)\big|_{X_1=0} ~\times~\rho_1'(1-x_1^{(\ell)})\big|_{x_1^{(\ell)}=0},
\end{align*}
and \vspace*{-1em}
\begin{align*}
	\lambda_1'(X_1,X_2)\big|_{X_1=0} 	&= \sum_{j\geq 1} \lambda_{[2~j]}~(x_0)^j ,\\
	\rho_1'(1-x_1^{(\ell)})\big|_{x_1^{(\ell)}=0}  	&=\rho_1'(1).
\end{align*}
Finally, we derive the stability condition as follows:  
\begin{align*}
	\lambda_1'(x_1^{(\ell)},x_0) \times \rho_1'(1-x_1^{(\ell)})\big|_{x_1^{(\ell)}=0} &= \sum_{j\geq 1} \lambda_{[2~j]}~(x_0)^j \times\rho_1'(1)< 1.
\end{align*}

\section{Proof of Theorem~\ref{Stability_Raptor_tandem}} \label{appendix_C}
We use the following notation throughout: 
$a_2^{(\ell)}$ and $b_2^{(\ell)}$ respectively denote the PDFs of  variable-to-check message and check-to-variable message along 
$\Pi_2$  at the $\ell$th BP decoding iteration, $\tilde{x}^{(\ell)}$ and $\tilde{y}^{(\ell)}$ respectively denote
the D-means~{\protect\cite[pages 201]{richardsonBOOK2008modern}} associated with $a_2^{(\ell)}$ and $b_2^{(\ell)}$, and 
$a_0$ and $\tilde{x_0}$ respectively denote the PDF of channel LLR and the D-mean associated with $a_0$.

Using equations ~(\ref{eq:VN to CN PDF MET}) and~(\ref{eq:CN to VN PDF MET}), 
we can write  DE equations related to variable nodes and check nodes for the LT part of the Raptor code represented in the multi-edge framework as follows:
\begin{align}
a_2^{(\ell+1)} &=  \sum_{i\geq1}  \lambda_{i}\left[b_2^{(\ell)}\right]^{\otimes (i-1)}  = \lambda_2(b_2^{(\ell)}), \label{t1} \\
b_2^{(\ell)} &= a_0 \boxtimes \sum_{i\geq 0} \rho_i\left[a_2^{(\ell)}\right]^{\boxtimes (i-1)} 
= a_0 \boxtimes \rho_2(a_2^{(\ell)}), \label{t2}
\end{align}
where $\lambda_2(x)$ and $\rho_2(x)$ is given in (\ref{tandem_lamda}) and (\ref{tandem_rho}). 
Then applying Lemma~4.60 given in~{\protect\cite[pages 202]{richardsonBOOK2008modern}} to  (\ref{t1}) and (\ref{t2}) give 
\begin{align*}
\tilde{y}^{(\ell)} =  \tilde{x_0}~ \rho_2(\tilde{x}^{(\ell)}) \hspace*{1cm} \text{and} \hspace*{1cm}
\tilde{x}^{(\ell+1)} \leq 1-\lambda_2(1-\tilde{y}^{(\ell)}).
\end{align*}
This finally gives us the update rule for 	$\tilde{x}^{(\ell+1)}$ as 
$\tilde{x}^{(\ell+1)} \leq 1-\lambda_2\left(1-\tilde{x_0} ~\rho_2(\tilde{x}^{(\ell)})\right)$.
Furthermore, to ensure that the BER decreases throughout the BP decoding iterations condition, $\tilde{x}^{(\ell+1)}\geq \tilde{x}^{(\ell)}$ must hold. Thus, for successful starting under DE, we need to guarantee that 
\begin{align}
\tilde{x} &\leq 1-\lambda_2\left(1-\tilde{x_0} ~\rho_2(\tilde{x})\right). \label{t_3}
\end{align}
For the decoding to successfully start, inequality (\ref{t_3}) needs to be valid around around $\tilde{x}=0$. Therefore, the derivative of the left-hand side is majorized by the derivative of the right-hand side at zero, which shows that
\begin{align}
\left(\frac{2 \alpha}{\beta}~\Omega_2 ~\tilde{x_0}\right) e^{\left(-\frac{\alpha}{\beta} \Omega_1 \tilde{x_0}\right)} \geq 1, \label{t_4}
\end{align}
where $\alpha$ and $\beta$ respectively denote the average input bit node degree and average output bit node degree.
Note that we can follow the same procedure as given in Appendix~\ref{appendix_B} to compute the derivative of (\ref{t_3}). Finally (\ref{t_4}) gives 
\begin{align*}
\Omega_1 \geq 0 \hspace*{1cm} \text{and} \hspace*{1cm}
\Omega_2 \geq \frac{\beta}{2 \alpha \tilde{x_0}} .
\end{align*}
Note that the ratio between $\beta$ and $\alpha$ gives the rate of the Raptor code,	$\mathcal{R}(\gamma)$, at SNR  $\gamma$. And if a code is capacity achieving, then $\mathcal{R}(\gamma) \approx  C(\gamma)$, where $C(\gamma)$ is the capacity of the channel at SNR  $\gamma$.  Therefore, for a capacity approaching Raptor code with degree distribution $\Omega(x)$, we have
$\Omega_2 \geq {C(\gamma)}/{2  \tilde{x_0}}$.

\end{document}